\documentclass{wiley-article}

\usepackage{siunitx}
\usepackage{caption} 
\usepackage{booktabs}
\usepackage[inline]{enumitem}
\usepackage{multirow}
\usepackage{lscape}
\usepackage{pifont}
\usepackage{xurl}
 \usepackage{setspace, amsmath, amssymb, url, lscape, subfig, algorithmic, multirow, pslatex, listings, verbatim, alltt, amsfonts, wrapfig, boxedminipage, color, cite, bookmark}
\usepackage{algorithmic}
\usepackage{algorithm}
\usepackage{balance}
\usepackage{nicematrix}
 
\papertype{Survey Article}
\paperfield{Journal Section}

\newcommand{\eg}{{\it e.g., }}
\newcommand{\etal}{{\it et~al. }}
\newcommand{\ie}{{\it i.e., }}

\newcommand{\comments}[1]{}
\newcommand{\cmark}{\ding{51}}
\newcommand{\xmark}{\ding{55}}

\newcommand{\revisedagain}[1]{#1}
\newcommand{\finalrevise}[1]{#1}

\DeclareMathDelimiter{(}{\mathopen} {operators}{"28}{largesymbols}{"00}
\DeclareMathDelimiter{)}{\mathclose}{operators}{"29}{largesymbols}{"01}
\newcolumntype{C}[1]{>{\centering\arraybackslash}m{#1\textwidth}}

\title{\revisedagain{Efficiency in the Serverless Cloud Paradigm: A Survey on
the Reusing and Approximation Aspects}}

\begin{document}
\author[\authfn{1}]{Chavit Denninnart}
\author[\authfn{1}]{Thanawat Chanikaphon}
\author[\authfn{1}]{Mohsen Amini Salehi}

\contrib[\authfn{1}]{Equally contributing authors.}

\affil[1]{High Performance Cloud Computing (HPCC) Laboratory, School of Computing and Informatics, University of Louisiana at Lafayette, Louisiana, 70503, USA}

\corraddress{Mohsen Amini Salehi, High Performance Cloud Computing (HPCC) Laboratory, School of Computing and Informatics, University of Louisiana at Lafayette, Louisiana, 70503, USA}
\corremail{amini@louisiana.edu}


\fundinginfo{\finalrevise{CNS-2007209, CNS-2007209}}

\runningauthor{Chavit Denninnart, Thanawat Chanikaphon, and Mohsen Amini Salehi}

\begin{frontmatter}
\maketitle

\begin{abstract}
Serverless computing along with Function-as-a-Service (FaaS) is forming a new computing paradigm that is anticipated to found the next generation of cloud systems. The popularity of this paradigm is due to offering a highly transparent infrastructure that enables user applications to scale in the granularity of their functions. Since these often small and single-purpose functions are managed on shared computing resources behind the scene, a great potential for computational reuse and approximate computing emerges that if unleashed, can remarkably improve the efficiency of serverless cloud systems---both from the user's QoS and system's (energy consumption and incurred cost) perspectives. Accordingly, the goal of this survey study is to, first, unfold the internal mechanics of serverless computing and, second, explore the scope for efficiency within this paradigm via studying function reuse and approximation approaches and discussing the pros and cons of each one. Next, we outline potential future research directions within this paradigm that can either unlock new use cases or make the paradigm more efficient.
\end{abstract}


\end{frontmatter}

\section{Introduction}
\label{sec:intro}

\subsection{Serverless Computing Paradigm}
The first generation of cloud technology, established around 2010, mitigated the burden of system administration and maintenance via consolidating servers and forming centralized data centers. It is anticipated that the second generation of cloud technology focuses on mitigating the burden of developing cloud-native applications for programmers and solution architects via the serverless computing paradigm \cite{schleier2021serverless}.

Serverless computing provides the developers with high-level software abstractions, such as functions, a.k.a. Function-as-a-Service (FaaS), and transparently deploying them, such that the user has the illusion of having no servers to  manage~\cite{lopez2021serverless}. Accordingly, modern software engineering methodologies, such as DevOps \cite{ebert2016devops} and Continuous Integration Continuous Delivery (CI/CD) pipelines \cite{arachchi2018continuous}, have adopted the serverless computing paradigm to facilitate rapid cloud-native application development. These methodologies instruct splitting an application into several functions that are invoked periodically or in response to an event. Behind the scene, each function invocation leads to the execution of one or an ordered set of stateless microservice(s)~\cite{lloyd2018serverless}. 

\begin{figure}[ht]
    \centering
    \includegraphics[width=0.55\textwidth]{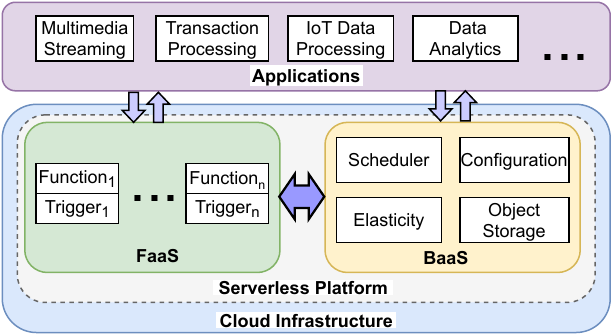}
    
    \caption{\small{The serverless computing paradigm mitigates the burden of developing cloud-native applications via offering a high-level programming abstraction (FaaS) and transparently executing them (BaaS). Different applications can define and trigger functions with minimal configurations needed for each function.}
    }
    \label{fig:gpserverless}
\end{figure}

As shown in Figure~\ref{fig:gpserverless}, the serverless computing paradigm can be defined as the combination of FaaS and BaaS (Backend-as-a-Service) subsystems (\ie Serverless = FaaS + BaaS~\cite{schleier2021serverless}). While FaaS focuses on the front-end development of functions in a wide variety of programming languages, BaaS focuses on the transparent and isolated execution of the functions. BaaS is also in charge of data storage, scheduling, monitoring, and transparent elasticity of the functions.
It is noteworthy that serverless computing is a loose term, and it does not strictly enforce the user's code to be based on FaaS. Moreover, Serverless solutions are sometimes abstracted from the user perspective. For instance, Amazon Athena~\cite{awsathena} is an interactive SQL-like query processing service for Amazon S3 data. Although Athena operates based on serverless principles, its users may consider it as a Platform-as-a-Service (PaaS) instead.

A common approach to handle function calls (henceforth, called \emph{users' requests} or \emph{tasks}) in BaaS is to gather the requests from all triggering sources (\eg API calls, timer, and events) into a central queue. Then, a resource allocator maps these requests to scalable pools of computing resources. To isolate the requests from each other and to avoid the side effects of using shared resources, often, some forms of task sandboxing, such as containers or micro Virtual Machines (\eg Firecracker VM \cite{agache2020firecracker}), are employed. 


\subsection{Scope for Efficiency in the Serverless Computing Paradigm}
The shared and transparent nature of serverless systems offers great potential for efficiency---both from the system and user perspectives. From the system end, metrics such as throughput, utilization, energy consumption, and carbon emission, and from the user end, Quality of Service (QoS) (\eg turnaround time), and the user's incurred cost can be potentially improved. 

The potential efficiency improvement can be unleashed, primarily via smart resource allocation methods that can identify identical and/or similar tasks in the serverless system. As a motivating example, consider the case of a serverless cloud used for processing live video contents before streaming them to the viewers~\cite{li2018cost}. As shown in Figure~\ref{fig:mergeapprox}, the system has \texttt{transcode(v,c)} function to change the codec of video segment \texttt{v} to \texttt{c}; and  \texttt{bitrate(v,b)} function to change the bit-rate of video segment \texttt{v} to \texttt{b}. The figure shows possible scenarios of function execution in the system. Consider two invocations of  \texttt{transcode(v1,c1)} and \texttt{transcode(v1,c2)} coexist in the system. Without merging, shown in Figure~\ref{fig:mergeapprox}(a), the two invocations separately \texttt{load}, \texttt{decode}, and \texttt{encode} the video. Alternatively, by merging these invocations into one task, shown in Figure~\ref{fig:mergeapprox}(b), the \texttt{load} and \texttt{decode} identical operations can be reused, and then \texttt{encode} operation into two different codecs is carried out individually.  Because \texttt{transcode()} function cannot be approximated, consider \texttt{bitrate()} function to explain function approximation. In this case, shown in Figure~\ref{fig:mergeapprox}(c), one invocation, \texttt{bitrate(v1,b2)}, can be approximated to \texttt{bitrate(v1,b1)}, hence, the whole execution chain can be reused. Accordingly, two main directions to improve the efficiency of serverless computing can be enumerated as follows: 

\begin{figure}
    \centering
    \includegraphics[width=0.6\textwidth]{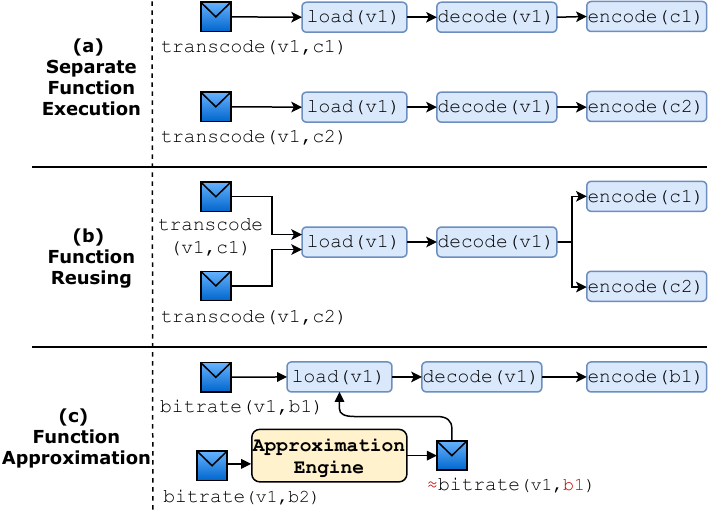}
    
    \caption{\small{Comparison of three scenarios to handle similar function invocations: (a) separately executing functions; (b) reusing \texttt{load} and \texttt{decode} microservices of the function calls, and (c) approximating a function call and making it compatible with an existing one.
    }}
    \label{fig:mergeapprox}
\end{figure}

\begin{enumerate}[]
    \item \emph{Computational reuse} that avoids redundant processing of identical or similar function requests. It focuses on reusing the whole or part of the execution, underlying platform (\eg container), and allocated resources of a process. A well-established reusing approach is based on caching \cite{carlson2013redis} that can avoid the re-execution of a recent task. While caching is \emph{retroactive} by nature and can only capture identical tasks, in serverless computing, there is a scope for \emph{proactive} reusing. In this manner, similar (or identical) concurrent function calls can be aggregated to one merged task to reuse a part of (or the whole) computation---even before an instance of the task is complete and cached. For example, the scheduler of a serverless system can detect two workflows that share the same sub-task (or use the same data) ahead of time, and schedule the sub-task or data to be reused. 
    
    \item \emph{Approximate computing} that can be employed in contexts where lower quality (less accurate) results can be tolerated (\eg Machine Learning (ML), live-streaming, etc.). Using approximate computing, the cost, energy, and response time of the serverless cloud can be reduced. Some common approaches for approximate computing include scaling down the precision of the invoked function, downsampling its input data, skipping some computation steps in the function workflow, or approximating the function result from similar or recent invocations. 
\end{enumerate}

It is noteworthy that, in use cases where security is a concern, function reuse and approximation can be carried out on subsequent function invocations of the same user. Nevertheless, in other use cases where there is no security concern (\eg stateless functions performing mathematical operations \cite{wis23}) reusing and approximation can be implemented more broadly---across users or organizations.

\subsection{Positioning of This Survey Study}

To position this survey paper, we describe recent related studies from academia and industry that focus on serverless computing, and then, in Table \ref{tab:serverlesssurveys}, we summarize the comparison and highlight the topics covered in each work.

\begin{enumerate}[label=(\roman*), wide, labelwidth=!, labelindent=0pt]
\item Optimizing and extending serverless platforms by Nazari \etal \cite{nazari2021optimizing}: This study focuses on the cold start and startup time optimizations, inter-function communication techniques, and possible extensions to the serverless platform. Although the serverless optimization aspects overlap with our work, we concentrate on the reusing and approximate computing techniques that are not covered in them.
\item Architectural design of serverless systems by Li \etal \cite{li2021serverless}: This work decouples the serverless architecture into four stacked layers namely, virtualization, encapsulation, orchestration, and coordination. The survey includes multiple techniques of implementation and efficiency improvements, such as pre-warming and scheduling strategies. In contrast, our work encompasses optimization techniques that are achieved at the intersection of the platform and application levels.
\item Eismann \etal \cite{eismann2021state} survey the state of serverless applications by providing a systematic study of serverless applications. They analyze 16 characteristics of 89 serverless applications collected from open-source projects and literature and compared the results with 10 related survey studies and datasets to analyze community consensus.
\item Raza \etal \cite{raza2021function} measure various features and performance metrics of the commercial and open-source FaaS platforms. They particularly study the performance per cost benefits from the application developer’s perspective. Several optimization techniques are discussed to present a developer's decision-making factor in choosing a FaaS platform.
\item \revisedagain{Serverless computing survey conducted by Mampage \etal \cite{mampage2022holistic}: This work identifies aspects of serverless resource management and proposes a taxonomy of elements that influence these aspects, encompassing characteristics of system design, workload attributes, and stakeholder expectations. However, the proposed taxonomy does not include efficiency improvement aspects such as approximate computing.}
\item \revisedagain{The survey of opportunities and challenges in serverless by Li \etal \cite{li2022serverless}: This paper collects papers reflecting the state of the art of serverless computing. They identify the serverless computing model’s challenges and study how the existing works address them. Their work presents various areas that need further attention from the research community, on the contrary, our work particularizes on the scope for efficiency in the serverless paradigm.}
\item \revisedagain{Chakraborty \etal \cite{chakraborty2023journey} compare cloud frameworks on the basis of different paradigms. Cloud computing, fog computing, Cloud of Things (CoT), and Fog of things (FoT) paradigms are discussed with their respective properties and limitations. Our work is conducted with a comparable aspect, but dives deeper into the serverless paradigm.}
\item CNCF annual survey \cite{cncf22}: It is a statistical report that contains a list of up-to-date technologies and serverless providers. 
\item The state of serverless by Datadog \cite{datadog21}: The report analyzes the current trends of serverless applications on three leading cloud providers namely, AWS, Google Cloud, and Microsoft Azure.
\item Enterprise-level serverless systems by IBM \cite{ibm21}: This survey was conducted by IBM Market Development \& Insights (MD\&I) and encompasses the perception of enterprises from the serverless paradigm, in areas like user experience, security, and CEO opinions.
\end{enumerate}

\begin{table}[]
\centering
\resizebox{0.8\textwidth}{!}{
\begin{NiceTabular}{ccc|C{0.05}|C{0.05}|C{0.05}|C{0.05}|C{0.05}|C{0.05}|C{0.05}|C{0.05}|C{0.05}|C{0.05}|C{0.05}}[rules/color=[gray]{0.7},code-before = \rowcolor{gray!20}{7,8,9,10,11}]
 & \Block{1-2}{\textbf{\small{Prior studies}}} 
 & 
 & \textbf{\small{Naz. \etal \cite{nazari2021optimizing}}} 
 & \textbf{\small{Li \etal \cite{li2021serverless}}} 
 & \textbf{\small{Eism. \etal \cite{eismann2021state}}}
 & \textbf{\small{Raza \etal \cite{raza2021function}}} 
 & \textbf{\small{Mam. \etal \cite{mampage2022holistic}}} 
 & \textbf{\small{Li \etal \cite{li2022serverless}}} 
 & \textbf{\small{Chak. \etal \cite{li2022serverless}}}
 & \textbf{\small{CNCF \cite{cncf22}}} 
 & \textbf{\small{Data-dog \cite{datadog21}}} 
 & \textbf{\small{IBM \cite{ibm21}}} 
 & \textbf{\small{This survey}} 
 \\
\hline \hline
\Block[borders={right}]{5-1}{\rotate \small{Platform aspects}}
 & \Block[borders={right}]{3-1}{\small{Challenges}} & \small{stateful serverless}
 & \checkmark & \checkmark & & & \checkmark & \checkmark & & & & & \checkmark \\ \cline{3-14} 
& & \small{cold start}
 & \checkmark & \checkmark & & \checkmark & \checkmark & \checkmark & & & & & \checkmark \\ \cline{3-14}
& & \small{security}
 & & \checkmark & & & \checkmark & \checkmark & \checkmark & & & \checkmark & \checkmark \\ \cline{2-14}
& \Block[borders={right}]{2-1}{\small{Adoption}} & \small{trend}
 & & & \checkmark & \checkmark & & & \checkmark & \checkmark & \checkmark & \checkmark & \\ \cline{3-14}
& & \small{compare solutions}
 & & \checkmark & \checkmark & \checkmark & \checkmark & \checkmark & \checkmark & \checkmark & & & \checkmark \\
\hline\hline
\Block[borders={right}]{5-1}{\rotate \small{Performance aspects}}
 & \Block[borders={right}]{1-1}{\small{Reusing}} & \small{deterministic}
 & \checkmark & \checkmark & & \checkmark & \checkmark & \checkmark & & & & & \checkmark \\ \cline{2-14}
& \Block[borders={right}]{2-1}{\small{Approx.}} & \small{data level}
 & & & & & & & & & & & \checkmark \\ \cline{3-14}
& & \small{instruction level}
 & & & & & & & & & & & \checkmark \\ \cline{2-14}
& \Block[borders={right}]{1-1}{\small{Reuse \& approx.}} & \small{semantic}
 & & & & & & & & & & & \checkmark \\ \cline{2-14}
& \Block{1-2}{\small{Scheduling}} &
 & \checkmark & \checkmark & & \checkmark & \checkmark & \checkmark & \checkmark & & & & \checkmark \\
\hline \hline
& \Block{1-2}{\small{Future direction}} &
 & \checkmark & \checkmark & & \checkmark & \checkmark & \checkmark & \checkmark & & & & \checkmark
\end{NiceTabular}
}
\caption{\small{Positioning of this survey study with respect to prior survey studies in the serverless computing area.}}
\label{tab:serverlesssurveys}
\end{table}

A summary of the characteristics of these studies is compared in Table~\ref{tab:serverlesssurveys}. In the table, we can see that several prior works focus on the platform and specification aspects of the serverless systems, whereas, our work concentrates on the potential to improve the performance of serverless systems via novel computational reuse and approximate computing techniques. 

\subsection{Paper Structure}
Before studying efficiency in serverless cloud computing, we need to learn about the nuts and bolts of serverless computing. Accordingly, in the rest of this survey, we first dive deep into the serverless computing details and study its anatomy. Next, we compare the serverless systems against other distributed computing paradigms and discuss why a separate study is required to make these systems efficient. Then, we concentrate on the efficiency of the serverless systems. The examples described in the previous part only show one possible scenario for reusing and approximating a function. We are to explore the potential for different forms of function reuse and approximation that can be unleashed, thereby, enabling efficient serverless cloud computing. An overview of the approaches that are studied in this work is shown in Figure~\ref{fig:ReuseApprox}. 

We believe this survey study can help the research community to further develop these areas and build more efficient serverless computing platforms. 
The rest of the paper is organized as follows: 
Section~\ref{sec:serverlessIntro} introduces the current state of commercial and research-based serverless computing platforms. Section~\ref{sec:unique} address the unique characteristics of serverless that provide potential and burden to reusing and approximation techniques. Then, Section~\ref{sec:reuse} discusses the potential of computational reuse on various parts of the serverless computing platform. Next, Section~\ref{sec:approxcloud} discusses the approximate computing techniques that can be applied on serverless computing platforms. Section~\ref{sec:app} lists some potential development directions to increase data and compute reusability in serverless computing platforms. 
Finally, we conclude this paper in Section~\ref{sec:conclsn}.


\section{Nuts and Bolts of the Serverless Computing Paradigm}
\label{sec:serverlessIntro}

\subsection{Introducing Serverless Computing}
    Serverless computing and FaaS abstract the users from both server maintenance and management. However, unlike PaaS, using serverless entails breaking the application into multiple functions that each one can potentially be developed in a different programming language. Then, the entire function execution management, such as resource allocation, scaling, scheduling, fail-over, and platform configurations, are transparently handled by the underlying serverless platform. As such, this paradigm simplifies the software development process and enables the users to become solution-oriented and focus on their business logic, rather than specific server configuration details. 


\begin{figure*}
    \centering
    \includegraphics[width=0.45\textwidth]{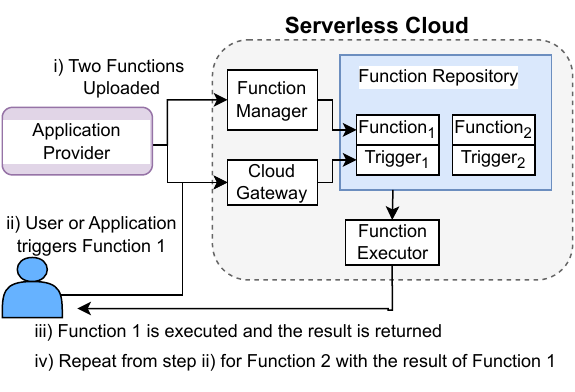}
    \includegraphics[width=0.45\textwidth]{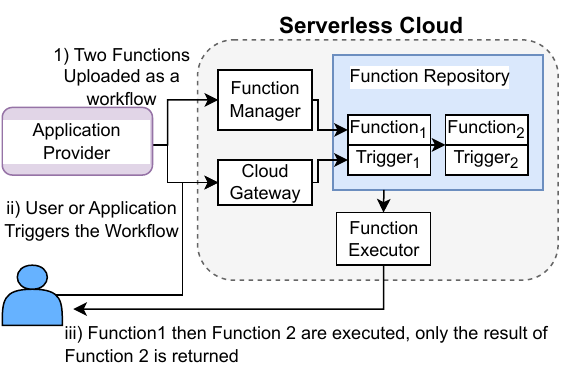}
    \caption{\finalrevise{An example workflow with two functions uploaded by the application provider. In this example, the Function trigger is configured as an on-demand API request and the result of Function 1 serves as the input for Function 2. Scenario A (on the left) sets up two functions separately. Scenario B (on the right) chains two functions on the serverless platform. }
    }
    \label{fig:Triggers}
\end{figure*}

\subsection{Functions Triggers in Serverless Computing}
\label{subsec:trigger}
FaaS enables users to develop functions and define a way to enact their execution, called  \textit{function triggers}. FaaS platforms desire the user's application to be constructed as a set of single-purpose functions that receive a set of input parameters and yield a set of outputs~\cite{scheuner2020function}. Each function is managed and scaled independently by the BaaS platform. 

Function triggers are typically based on API calls (\eg web requests), timers, and completion of other events (\eg completion of another task) \cite{peekserverless,shahrad2020serverless}.  
\finalrevise{As shown in Figure~\ref{fig:Triggers}, two main ways to designate function triggers for serverless workflows are: (a) Defining the trigger for each function individually; and (b) Utilizing a \emph{workflow} schema (\eg JSON workflow definition \cite{serverlessworkflow}). In scenario A of Figure~\ref{fig:Triggers}, the application provider creates two functions separately (through the function manager) in the serverless platform. Upon triggering the workflow through the cloud gateway, Function 1 is executed first. Then, the result is returned to the application provider, and the provider feeds the result of Function 1 to trigger Function 2 and get the intended result. In scenario B, however, the application provider chains two functions as a workflow on the cloud. In this case, upon Function 1 completion, the serverless platform automatically triggers Function 2 with the result of function 1 before returning the final result to the user.} 

Although defining an individual trigger for each function is simpler, \revisedagain{more flexible} and more popular, defining multiple function triggers together in form of a workflow schema is more advantageous. First and foremost, the schema can provide useful metadata for the resource allocator in BaaS to identify parallelizable tasks and schedule them together, thereby, improving the resource utilization and users' QoS (\eg waiting time).
Second, using the schema, complex function workflows can be defined that otherwise would be time-consuming and error-prone to build~\cite{burckhardt2021serverless}. The third advantage is the portability and reusability that making use of the workflow schema offers. While per-function triggers are tied to each function, the workflow of functions is defined within a schema file with a specific syntax~\cite{burckhardt2021serverless}. The schema can be used to effortlessly re-deploy the application on the same serverless platform. However, re-deploying the application on the different public serverless platforms is not achievable at this time, due to the lack of platform-agnostic standards.

\subsection{The Matter of ``Function State'' in Serverless Computing}
\label{subsec:serverlessstate}
Functions in serverless computing are originally designed to be stateless. That is, a function does not maintain (\ie memorize) any state data (\eg shared variables) between consecutive invocations, and its output is merely subject to its input arguments. Statelessness is, in fact, a primary practice in functional programming~\cite{bird1988functional} that prevents side effects~\cite{josephs1986functional}, thereby, improving software robustness and predictability. This implies that, for a given input, a stateless function (\eg mathematical operation, query/string preprocessing, etc.) always yields the same output, thus, the function results can be reused (\eg via caching). In addition, stateless functions mitigate the overhead of serverless platforms by relieving them from maintaining data consistency and synchronization in executing functions~\cite{lopez2021serverless}. 
 
Despite the benefits of stateless functions, some applications naturally demand the state to be maintained. Refactoring the stateless version of these applications makes them prohibitively inefficient. For instance, a big data analytics workload (\eg for semantic search \cite{zobaed2021senspick}) cannot afford to load the entire dataset for each function call, nor can it afford to forward the output to other functions along the workflow. A common approach to circumvent this situation is to persist the state on the external storage services \cite{jain2019spiitserve}. However, Pu~\etal~\cite{pu2019shuffling} demonstrate that employing external storage to carry out serverless data analytics is up to 500$\times$ slower than using IaaS clouds. Note that, once the state of a function is persisted on the external storage, it behaves as a stateful function and its results are not reusable anymore (unless the state domain is small and cacheable).

The matter of state is still an open challenge in the serverless paradigm. 
Several research works have been undertaken to offer a built-in stateful serverless solution~\cite{burckhardt2022netherite}. Such solutions often employ some forms of key-value and/or file-based storage. 
Sreekanti \etal~\cite{sreekanti2020cloudburst} develop a stateful serverless platform, called Cloudburst, using Anna~\cite{wu2020autoscaling}, which is auto-scaling key-value storage, to persist the state. Pu \etal propose Shuffling~\cite{pu2019shuffling}, a stateful domain-specific serverless platform for data analytics, with a hierarchical state persistence---a fast layer on the memory and a slower one on the device storage. 
Schleier-Smith~\etal develop a dedicated POSIX-like file-storage system to enable stateful serverless computing, called FAASFS~\cite{schleier2020faas}. It tackles multiple challenges of providing a shared file system across functions, such as cache and transactional consistency~\cite{yu2018sundial,sreekanti2020fault}. Shillaker and Pietzuch evaluate stateful functions within the FAASM platform~\cite{shillaker2020faasm} via sharing the state in form of both memory segments and files.
\revisedagain{Kraft \etal propose Apiary~\cite{kraft2022apiary}, a serverless framework for data-centric functions. It compiles application logic into a database stored procedures to improve performance.}
On the other hand, some solutions for the function state operate based on the actor model~\cite{hewitt2010actor}. Azure Functions provide support for the Entity functions~\cite{awsentityfunctions}. Kalix~\cite{akkaserverless} and Apache Flink~\cite{apacheflink} offer an implementation of the actor model for stateful serverless.


\subsection{Function Isolation in Serverless Computing}
\label{subsec:isolation}
\begin{figure}[ht]
    \centering
    \includegraphics[width=.7\textwidth]{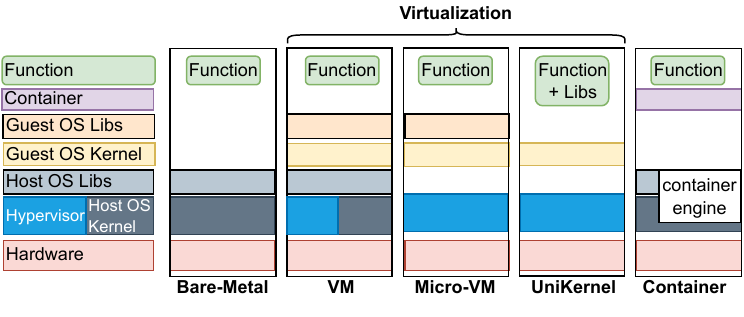}
    
    \caption{\small{A bird-eye view of the underlying layers of various isolation platforms for serverless functions. From left to right, respectively, there are functions on bare-metal (\eg via WebAssembly), various forms of VMs, and a container. The number of layers implies the overhead of each isolation platform.} 
    }
    \label{fig:isolationArchitecture}
\end{figure}

In principle, virtualization is not a must for FaaS and serverless cloud offerings. A user can essentially call a function using a command in the general form of \texttt{client.invoke(FunctionName=`F',Payload=Data)} \cite{mcgrath2017serverless,perez2018serverless}. Upon invocation, the FaaS engine can interpret the function and form a task that can be then directly executed on the host machine (\ie bare-metal resource provisioning). However, lack of isolation in bare-metal raises security concerns, particularly, when there are coexisting tasks from multiple users on shared computing resources. Therefore, some form of sandboxing is required to isolate the execution environment of each function call. Broadly speaking, such isolation can be provided at the following levels: application-level runtime frameworks (WebAssembly \cite{kjorveziroski2022evaluating} and language runtime \cite{carreira2021warm}), Operating system-level (containerization), and hardware-level (virtualization). The layer-view of each isolation platform for the serverless functions is provided in Figure~\ref{fig:isolationArchitecture} \cite{davoodpaper22}. The software stack of each platform implies the overhead imposed by that platform. In this figure, the left-most boxes serve as the legend---dedicating a color for each layer. The white space(s) in each isolation platform express the absence of the corresponding layer(s), represented on the left-most side. \revisedagain{For example, On the Hypervisor (blue color) and Host OS Kernel (dark gray color) level, the existence of these two colors on the VM column \finalrevise{implies} that both of these layers exist in the VM technology. Similarly, for Micro-VM and Uni-Kernel, it is blue only (without gray) meaning that only the Hypervisor exists (without Host OS Kernel).}

\textbf{WebAssembly:}
WebAssembly (a.k.a. Wasm) is an open standard that
enables the generation of portable binary code from various high-level programming languages and interfaces the binary code with the underlying host environment. The binary can be executed both as a standalone code or within a Web browser. WebAssembly and, particularly, its software-fault isolation (SFI) feature \cite{shillaker2020faasm} provides software-level isolation that can be used by serverless function solutions. FAASM~\cite{shillaker2020faasm} is an instance of a serverless framework that executes functions on WebAssembly. Also, in Kruslet~\cite{krustlet},  containers are replaced with WebAssembly in the context of the Kubernetes orchestrator \cite{sayfan2017mastering}. Kruslet listens to the Kubernetes event stream and upon receiving a task request, it executes the task on WebAssembly runtime~\cite{lopez2021serverless}, instead of creating (container) pods. 

\finalrevise{Using Wasm in a WebAssembly-based serverless platform eases function migration across devices, edge, and cloud. Since Wasm runtime is available on web browsers, the same packaged serverless function not only can be executed on the cloud server, but also on the end-users' web browsers. This allows the web application to have an adaptive function scheduling such that in a certain scenario (\eg user device has low battery or limited computing power), the function is executed on the cloud, whereas, in other scenarios (\eg low bandwidth) the function is executed locally on the device.}
 
Although WebAssembly can satisfy the needs of small-size functions, in many use cases, a specific environmental setup, such as software packages and libraries, is required. Just-in-time preparation of these dependencies for each function execution is time-consuming. Moreover, making use of WebAssembly implies compiling the functions to WebAssembly which curbs the generality of the serverless solution. Therefore, for the sake of generality and to maintain cost- and time-efficient preparation of functions, container or lightweight virtual machine (VM) technologies are more commonly used in the serverless domain.  

\textbf{Language runtime:}
\revisedagain{Similar to using WebAssembly, the language runtime can also be used as a lightweight execution environment. Dukic \etal~\cite{dukic2020photons} demonstrate overhead reduction via co-locating concurrent instances of the same function within the same runtime. However, it limits supported language and requires application modifications. Bruno \etal~\cite{bruno2022graalvisor} develop a virtualized polyglot language runtime for serverless applications, called Graalvisor, using Truffle~\cite{wimmer2012truffle}, an interpreter-writing framework, to relax language limitations.}

\textbf{Containerization:}
The most common way to package and isolate the function is through containerization~\cite{spe22song}. In this technology, a function is encapsulated within a widely-accepted container standard, termed Open Container Initiative (OCI) format~\cite{ociformat}, which is supported in all modern containerization solutions. Any programming language and/or software dependency can be supported, thus, the desired generality of serverless is accomplished. Unlike VMs that emulate the entire operating system stack, containers share the host kernel, thereby, 
both the memory and storage footprints are reduced. This pattern of reusing the kernel is extended to other layers within the container image. Specifically, container images have a layered structure that encourages reusing software packages across these images. Container engines use a method, called union mounting, through which a container is formed dynamically (\ie on-demand) via fetching its (read-only) layers at the runtime. Further details about union mounting are discussed in Section~\ref{subsec:reuse_process}. 

\textbf{Virtualization:}
Virtualization is the traditional method of providing strong isolation in cloud computing \cite{aminispe14}.
Due to including the whole operating system and application stack in the VM image, in general, VMs suffer from high memory and storage footprints. In addition, VMs introduce a high startup delay~\cite{ghatrehsamani2020art} to boot up, hence, they are not a perfect fit for frequent starts and terminations inherent to the serverless functions~\cite{shahrad2020serverless}. As such, VMs are usually employed as the underlying platform of the serverless frameworks, rather than an isolator of each function. In this case, function isolation within the VM is offered either via application-level solutions (\eg WebAssembly) or containers.  

\textbf{Micro-VM:} Although VMs are generally not ideal isolators for functions, there are some notable efforts to customize VMs for functions.
Firecracker~\cite{agache2020firecracker} is an AWS open-source project that provides a lightweight VM (a.k.a. micro-VM) with high isolation and low startup delay. It is being used by the AWS FaaS and serverless platforms (\eg AWS Lambda and AWS Fargate). Firecracker works similar to other full VM technologies that offer an isolated operating system environment to the user. However, unlike other KVM-based VMs~\cite{deepak} that sit in the user space on top of QEMU~\cite{bellard2005qemu} (a machine emulator that enables VMs), Firecracker directly communicates with the KVM layer via a customized emulation stack. While such a highly simplified emulation stack is sufficient for ordinary tasks, it lacks some notable features, such as libraries to support GPU and specialized CPU instructions. \revisedagain{As a result, serverless platforms that use Firecracker, such as AWS Lambda, are currently unable to provide services demanding advanced facilities, e.g., those needed for GPU-based machine learning algorithms~\cite{spe21}.}
Another downside of Firecracker is its deployment inflexibility because it only has to be deployed on top of the KVM hypervisor.

\textbf{Unikernel} \cite{gain2022want} is another lightweight VM-based technology. Similar to micro-VMs \revisedagain{(see Table~\ref{tab:virtualization} which compares all the aforementioned  isolation techniques)}, it bypasses the user space of the hypervisor. Moreover, it bypasses the user space of the guest operating system too. Thus, to run a function inside Unikernel, required guest OS libraries have to be incorporated at the application level. That is, the applications that are deployed within Unikernel have to encapsulate all their required libraries. Storing libraries on each function image causes substantial data redundancy and overhead. As such, Unikernel utilization becomes limited to small functions that do not require large dependencies. 

\begin{table}[]

\resizebox{\textwidth}{!}{
    \small
    \begin{NiceTabular}{c||c|c|c|c|c}%
        [code-before = \rowcolor{gray!20}{3,5,7}]
          & \textbf{Memory Cost} & \textbf{Startup Delay} & \textbf{Security} & \textbf{Main Pro} & \textbf{Main Limitation}  \\ \hline  \hline

         \textbf{WebAssembly} &	Low	& Short	& Software-fault Isolation & Wasm  runtime available on web browsers & Limited language and library support\\ \hline
         
         \Block{1-1}{\textbf{Language} \\ \textbf{runtime}} &	Low	& Short	& Software-fault Isolation & Ultra lean for an application-level isolation & Limited language and library support\\ \hline
        
        \textbf{Containerization} &	Low	& Short	& Containerization & Built-in container image partial reusing & Container platform security threats \\ \hline
        
        \textbf{VM} & High & Long	& VM Memory Isolation & Accelerator access; Proven isolation & High startup overhead \\ \hline
        
         \textbf{Micro-VM} &	Low	& Short	& VM Memory Isolation & Ultra lean for a VM-based isolation & \Block{1-1}{*Unsuitable for task that \\ require large software dependencies}   \\ \hline
         
         \textbf{Unikernel} & Low to Medium	* & Depends * & VM Memory Isolation & Better compatibility than Micro-VM & \Block{1-1}{ * Stores library within the function image, \\ thus, memory footprint may increase}
         \\
    \end{NiceTabular}
}
    \caption{\small{\revisedagain{Comparison of isolation techniques with respect to their memory cost, startup latency, security, and their main pros and limitations for serverless platform.} }}
    \label{tab:virtualization}
\end{table}

\subsection{Cold Start vs Warm Start Functions}

In a container-based serverless platform, a functioning container that resides in the memory to be launched rapidly is generally referred to as a \emph{warm start} container. In contrast, a function that must be loaded from the storage system in an on-demand manner is referred to as a \emph{cold start} container~\cite{lloyd2018serverless}. The cold start function involves loading the container image that (depending on the container size) imposes a nontrivial time overhead and can potentially dominate the function execution time \cite{lloyd2018serverless}. The overall cold start overhead of a function can be approximated by multiplying the function invocation frequency and the cold start overhead. 

Note that, calculating the cold start overhead can be further complicated when other system factors, such as elasticity and storage location, are taken into consideration. Importantly, there can be a lower-level cold start in which a function has to undergo the elasticity overhead and wait for the underlying VM (or hardware) to be made available before it can be loaded into its memory. Depending on the storage location, the cold start overhead can be subdivided into multiple tiers of cold, namely \emph{local storage cold} and \emph{repository cold}. In the former, the function container can be retrieved from the local storage, whereas, in the latter, the function must be retrieved from remote storage (\eg on a central cloud), which implies a more substantial overhead. These factors show that in an efficient serverless system, the BaaS subsystem must handle the cold/warm start of each function based on its characteristics. In Section~\ref{subsec:contreuse}, we discuss multiple approaches to strategically manage the containers and mitigate the cold start frequency.

\subsection{Serverless Cloud Solutions}
In this part, we first survey various commercial and open-source serverless cloud systems, and then, compare them (in Table~\ref{tab:serverless_platformcmp}) based on the aspects described in the previous sections. We note that, in addition to the platforms listed in Table~\ref{tab:serverless_platformcmp}, there have been several other serverless computing projects (\eg Fission Workflow \cite{fissionflow}, Kubeless \cite{kubeless}, and Iron Function \cite{ironio}) that were discontinued, thus, we have excluded them from the comparison table. 
\revisedagain{Because programming languages have an impact on function efficiency \cite{jackson2018investigation, bortolini2019investigating}, in Table~\ref{tab:serverless_platformcmp}, we include the list of language runtimes that are officially supported by each serverless system to help function developers in choosing a feasible system and language for their function.}
Lastly, in Section~\ref{subsec:generalarch}, we leverage our observations from the studied serverless systems and design a generic architecture  that includes the main components of the serverless systems.

\begin{table}[]
\centering
\resizebox{0.7\textwidth}{!}{
    \small
    \begin{NiceTabular}{c|c|c|c|c|c|c|c}%
    [code-before = \rowcolor{gray!20}{3,5,7,9,11,13,15}]
        \Block{1-1}{\rotate \textbf{Serverless} \\ \textbf{System}}
         & \Block{1-1}{\rotate \textbf{Open-Source} }
         & \Block{1-1}{\rotate \textbf{Container} \\ \textbf{Feed Format}}
         & \Block{1-1}{\rotate \textbf{Supported} \\ \textbf{Programming} \\ \textbf{Languages in} \\ \textbf{Function} \\ \textbf{Feed Format}}
         & \Block{1-1}{\rotate \textbf{Underlying} \\ \textbf{Platform}}
         & \Block{1-1}{\rotate \textbf{Stateful}}
         & \Block{1-1}{\rotate \textbf{Function} \\ \textbf{Start}}
         & \Block{1-1}{\rotate \textbf{Workflow} \\ \textbf{Support}}
         \\
        \hline \hline
        AWS Lambda &  \xmark    &  \cmark    & \small{\Block{1-1}{JS, Python, Go, Java,\\ Ruby, .NET, PowerShell}}  & \small{Firecracker}
         & \xmark & \small{\Block{1-1}{Cold/\\Warm}}
         & \cmark 
         \\

        \hline
        AWS Fargate & \xmark   &  \cmark    & \small{\Block{1-1}{Function Feed Format\\ not supported }} & \small{\Block{1-1}{Container on \\Firecracker}}
        
        & \cmark & \small{\Block{1-1}{Warm}} & \cmark 
        \\
        
        \hline
        \Block{1-1}{Microsoft \\ Azure Functions} &  \xmark   &  \cmark    & \small{\Block{1-1}{JS, Python, Java,\\ .NET, PowerShell}}  & \small{\Block{1-1}{Kubernetes\\/Azure Arc}}
        & \xmark & \small{\Block{1-1}{Cold/\\Warm}}
        & \cmark 
        \\
        \hline
        \Block{1-1}{Microsoft \\ Durable Function} &  \xmark   &  \cmark    & \small{\Block{1-1}{JS, Python, Java,\\ .NET, PowerShell}}  & \small{\Block{1-1}{Kubernetes\\/Azure Arc}}
         & \cmark & \small{\Block{1-1}{Warm}}
        & \cmark 
        \\
        
        \hline
        \Block{1-1}{ Google \\ Cloud Functions} & \xmark   &  \xmark    & \small{\Block{1-1}{JS, Python, Go, Java,\\ PHP, Ruby, .NET}} & \small{\Block{1-1}{Google\\App Engine}}
        & \xmark & \small{\Block{1-1}{Cold/\\Warm}} & \xmark \\

        \hline
        \Block{1-1}{ Google \\ Cloud Run} & \xmark   &  \cmark    & \small{\Block{1-1}{JS, Python, Go, Java,\\ PHP, Ruby, .NET, Kotlin}} & \small{\Block{1-1}{Kubernetes}}
        & \cmark & \small{\Block{1-1}{Cold/\\Warm}} & \cmark  \\
        
        \hline
        \Block{1-1}{ IBM \\ Cloud Functions} & \xmark   &  \cmark    & \small{\Block{1-1}{JS, Python, Go, Java,\\ PHP, Ruby, .NET,}} & \small{OpenWhisk}
         & \xmark &  \small{\Block{1-1}{Cold/\\Warm}} & \xmark  \\
  
  \hline

        OpenFaaS &  \cmark    &  \cmark    & \small{\Block{1-1}{JS, Python, Go, Java,\\ PHP, Ruby, .NET,}} & \small{\Block{1-1}{Kubernetes/\\OpenShift}}
          & \xmark & \small{\Block{1-1}{Cold/\\Warm}} & \xmark  \\ 
         
        \hline
        \Block{1-1}{Apache \\ OpenWhisk}   & \cmark    &  \cmark    & \small{\Block{1-1}{JS, Python, Go, Java,\\ PHP, Ruby, .NET,}} & \small{\Block{1-1}{Kubernetes/\\OpenShift/\\Docker}}
         & \xmark & \small{\Block{1-1}{Cold/\\Warm}} & \xmark  \\  

          \hline
        \Block{1-1}{Platform9 \\Fission}   & \cmark   &  \xmark    & \small{\Block{1-1}{JS, Python, Go, Java, PHP,\\ Ruby, .NET, Rust, Swift }}  & \small{Kubernetes} & \xmark & \Block{1-1}{Cold/\\Warm} & \xmark \\ 
         \hline

         Oracle Fn & \xmark  &  \cmark    & \small{\Block{1-1}{JS, Python, Go, Java,\\ Ruby, .NET}} & \small{Container}
         & \cmark & \small{\Block{1-1}{Cold/\\Warm}} & \cmark \\
      \hline
        

        Knative & \cmark   &  \cmark    & \small{\Block{1-1}{JS, Python, Go,\\ Java, Rust }}   & \small{\Block{1-1}{Kubernetes}}
        
        & \cmark & \small{\Block{1-1}{Cold/\\Warm}} & \cmark  
        \\     
        \hline
        
         Nuclio & \cmark   &  \cmark    & \small{\Block{1-1}{JS, Python, Go, Java,\\ .NET, Shell }}   & \small{\Block{1-1}{Kubernetes/\\Docker}}
        & \cmark & \small{\Block{1-1}{Cold/\\Warm}} & \cmark
                \\      
        \hline

        ƒuncX & \cmark   &  \xmark    & \small{\Block{1-1}{Python}}   & \small{\Block{1-1}{Kubernetes/\\Docker}} & \xmark & \small{\Block{1-1}{Cold/\\Warm}} & \xmark
                \\      
    \end{NiceTabular}
}
\caption{\small{Comparison of the major serverless computing platforms. }}
\label{tab:serverless_platformcmp}
\end{table}

\subsubsection{Public Serverless Cloud Platforms}
FaaS and serverless computing have commercially been made available via AWS Lambda service~\cite{awslambda} for the first time in 2014. AWS Lambda executes each function based on a user-defined trigger and charges the user only for the actual resource usage time (\ie the function execution time). The Lambda service arguably pioneered and shaped other FaaS services. Nowadays, Amazon also offers other serverless computing services---most notably AWS Step Functions \cite{awsstepfn} and AWS Fargate \cite{awsfargate}. AWS Step Functions is a workflow service that can chain a sequence of Lambda functions and other AWS services to build a serverless application. It manages the workflow in terms of scheduling, failure, and parallelization so that the users can focus on the higher-level business logic. Alternatively, AWS Fargate operates based on containers rather than functions. It is an example of a serverless service that is not built from a FaaS platform.


After AWS, multiple competing serverless computing cloud services, such as Azure Functions~\cite{msAzurefunctions}, Google Cloud Functions~\cite{gcloudfunctions}, and IBM Cloud Functions~\cite{ibmcloudfunctions} have emerged. These services are consistently evolving with different sets of features. Notably, recently, Microsoft Azure released Durable Functions \cite{msAzureDurablefunctions} to extend the Azure Function service to support stateful workflows.




\subsubsection{Private Serverless Cloud Platforms} 
Although public serverless platforms are increasingly popular, they come with the vendor lock-in risk and the trustworthiness issue that is inherent to public clouds. Therefore, multiple open-source projects have been developed to allow serverless deployment on self-hosted servers~\cite{Li2021Analyzing}.
\revisedagain{The details of stateful serverless platforms are described in Section~\ref{subsec:serverlessstate}.}
OpenFaaS~\cite{openfaas} and Apache OpenWhisk~\cite{openwhisk} are two popular open-source serverless platforms that dominate the private serverless cloud market.

OpenFaaS handles each function as a container that is deployed through Kubernetes. Therefore, a user can develop functions in the programming language of her choice. A packaging script creates a container image with the user's function encapsulated in it. Each function container is stored and managed in the Docker Registry \cite{dockerregistry} and also in the function store. While OpenFaaS is open-source and free to use, OpenFaaS PRO \cite{openfaas} is developed for commercial purposes.

OpenWhisk is another popular open-source serverless cloud platform backed by the Apache foundation \cite{Apache}. In comparison with OpenFaaS, OpenWhisk has a bigger developer community and many more features. Similar to OpenFaaS, the OpenWhisk project is also based on Kubernetes. It also utilizes many features from other open-source products, such as Kafka \cite{Kafka}, CouchDB \cite{CouchDB}, Nginx \cite{Nginx}, Redis \cite{carlson2013redis}, and Zookeeper \cite{Zookeeper}. This allows the OpenWhisk to be very scalable and, at the same time, feature-rich. However, this makes the learning curve of deploying and managing OpenWhisk steeper. OpenWhisk is also offered commercially on IBM Cloud Functions \cite{ibmcloudfunctions}. \revisedagain{Fission~\cite{fission} is another serverless platform heavily based on Kubernetes. However, in comparison with Kubeless, Fission has less dependency on Kubernetes as they implement their own function management components, whereas, Kubeless relies on Kubernetes components and extends only \finalrevise{the} required features that are missing to become a serverless platform.}


Fn Project~\cite{fnproject} is an open-source platform that works with Docker containers in its underlying layer. It supports API-based event triggers (\eg in form of web requests). Fn Flow \cite{fnflow} is an alternative version of Fn that can support workflows. Although both projects have attracted limited users from the open-source community, Oracle still offers them commercial serverless solutions. 

Knative~\cite{knative} is a fast-growing open-source project led by IBM and Google. Similar to Kubeless~\cite{kubeless}, which has been discontinued, Knative sits on top of Kubernetes and enables it to handle serverless workloads. According to the Cloud Native Computing Foundation (CNCF) survey \cite{cncf20}, Knative has been the top installable serverless solution in 2020. Knative project includes three main components, namely \emph{Build, Serve,} and \emph{Event}. Build is in charge of source code management, containerization, and making it deployable by Kubernetes. Serve deals with service deployment, managing microservice revisions, routing requests to different versions of microservices, automatic scaling, and scaling to zero. Finally, Event takes care of creating function triggers and forming workflow pipelines. 
\revisedagain{Knative is one of the backends supported by KNIX MicroFunctions \cite{knix} that is a well-known serverless platform in the academic area (formerly known as SAND~\cite{akkus2018sand}).}

\revisedagain{
Nuclio~\cite{nuclio} is another platform that works on top of Kubernetes. It offers popular data science tools integration and GPU-based machine support.
Lastly, the aforementioned solutions that leverage centralized management can potentially lead to some form of vendor lock-in. Hence, frameworks for decentralized function execution are proposed \cite{ciavotta2021dfaas,ghaemi2020chainfaas}. ƒuncX~\cite{li2022funcx} is another open-source FaaS platform for federated ecosystems. It supports function execution on remote heterogeneous computers and clouds.
}

\begin{figure}[ht]
    \centering
    \includegraphics[width=0.55\textwidth]{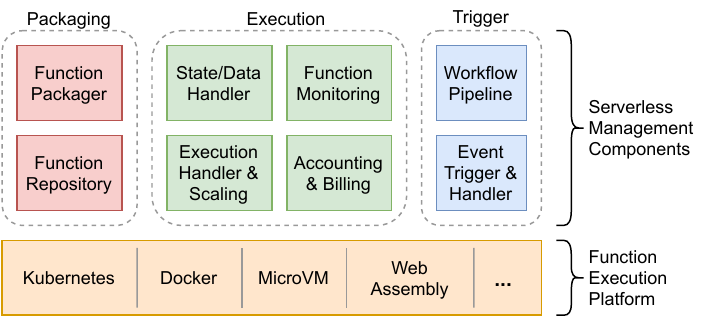}
    
    \caption{\small{Common architectural components of a serverless framework. Note that not all these components exist in every framework. For example, Function Packager exists only on the frameworks that do not run functions straight from the source code. The Function Execution Platform shows the list of common execution platforms for functions. However, this list is not inclusive, as there can be other execution platforms emerging or already deployed in the proprietary systems.
    }}
    \label{fig:commoncomponents}
\end{figure}

\subsubsection{General Architecture of Serverless Clouds}\label{subsec:generalarch}
Considering the common architectural components of the studied serverless platforms, we can design a generic architecture, depicted in Figure~\ref{fig:commoncomponents}, that is composed of an underlying function execution framework and eight components on top that take care of different aspects of serverless management. The ``Packaging'' component contains the Function Packager that wraps the function code into isolated units, such as container image, MicroVM image, and WASM compiled code \cite{shillaker2020faasm}. The ``Execution'' component is considered the engine of the serverless systems and is in charge of executing and monitoring functions. It also assures that the functions remain scalable and are accessible with low latency (via warm-starting functions). Particularly, it supports scale-to-zero when the function is unused. The State Handler module only exists to support stateful functions and is not used in stateless serverless platforms. The ``Trigger'' component on one end is used by the developer to declare the conditions for initiating a function call. On the other end, it connects with the Execution component to execute the task(s) based on the user requirements. The Workflow Pipeline is a high-level tool (\eg AWS Step functions \cite{awsstepfn}) that enables developers to create new services by defining a function chain.

Note that, a given serverless platform may only have a subset of these components, and it may categorize them sightly differently. For instance, Knative's main components (Build, Serve, and Event) can be mapped to \emph{Packaging, Execution,} and \emph{Trigger} group of components. Function Packager is only needed if the framework requires function code to be transformed into the container or other forms of the compiled function. Workflow Pipeline only exists in the frameworks that support function workflows.


\section{Serverless Computing VS Other Distributed \\
Computing Paradigms}

\label{sec:unique}



\begin{table}
    \centering
\resizebox{0.6\textwidth}{!}{
    \small
    \begin{NiceTabular}{c||c|c|c|c|c}%
    [code-before = \rowcolor{gray!20}{3,5,7,9}]
          & \Block{1-1}{\textbf{Task latency} \\ \textbf{sensitivity}} & \textbf{Request nature} & \textbf{Task types} & \Block{1-1}{\textbf{Approx. }\\\textbf{hardware} \\\textbf{accessibility}} & \Block{1-1}{\textbf{Stateless}} \\ \hline  \hline
         \textbf{HPC} &	Low	& Reservation	& User-defined &	Direct access &	No \\ \hline
         \textbf{Grid} &	Low	& Reservation	& User-defined &	Direct access &	No \\ \hline
\textbf{P2P} &	Can be high	& On demand task &	Predefined &	Direct access &	No \\ \hline
\textbf{IaaS} &	Low	& VM Lease	&  Predefined	& Virtualized	& No \\ \hline
\textbf{PaaS} &	Can be high	& On demand task &	\Block{1-1}{Predefined \\ types} &	Abstracted	& No \\ \hline
\textbf{SaaS} &	Generally high &	Interactive task &	Predefined &	Abstracted &	Can be \\ \hline
\Block{1-1}{\textbf{Edge-to-}\\\textbf{cloud}} &	Definitely high	& Interactive task &	\Block{1-1}{ Generally \\predefined} &	Abstracted 	& \Block{1-1}{Generally \\not} \\ \hline
\textbf{Serverless} &	Generally high &	\Block{1-1}{On demand/ \\Interactive task} &	User-defined &	\Block{1-1}{Abstracted \\Originally} & Yes 

    \end{NiceTabular}
}
    \caption{\small{Characteristics of the tasks and platforms across different distributed system paradigms.}}
    \label{tab:serverlessunique}
\end{table}

Many characteristics of serverless systems, including computational reuse and approximate computing, share similarities with other distributed computing paradigms, such as High-Performance Computing (HPC) systems, Grid computing, and various forms of cloud computing. However, the serverless paradigm exposes characteristics and obstacles that call for solutions specifically developed for them. In this section, we compare these various distributed computing paradigms (in Table~\ref{tab:serverlessunique}) and discuss those aspects of the serverless demanding new solutions.

Considering Table~\ref{tab:serverlessunique}, serverless tasks are usually user-defined  \revisedagain{(unlike pre-defined tasks in P2P, IaaS, PaaS and SaaS)}, are requested upon the user's demand \revisedagain{(as opposed to time/resource reservation or interactive task)}, and then are interactively served; as opposed to HPC systems that are reservation-based and offline. Unlike HPC and Grid computing paradigms, serverless tasks are often latency-sensitive. These characteristics entail having low-latency solutions to detect reusable tasks in the serverless system and acting upon them, whereas, existing solutions for HPC/Grid systems (\eg \cite{casas2017balanced, guo2018potluck, denninnart2019improving}) are designed for large offline tasks. Although the serverless functions are user-defined, they are stateless and fine-grained (\ie they are typically single-purpose), rather than complex stateful applications in other paradigms. These characteristics make the serverless systems potent for task duplication and reusing with proven performance gains \cite{denninnart2021harnessing, icsoc18}. \finalrevise{While serverless platforms can execute tasks in a pipeline to complete a workflow, each task in the workflow is often executed on the same unified execution engine. This is different from what commonly occurs across edge-to-cloud continuum, where the edge generally takes care of the pre-processing and the core processing happens on the cloud. }

\revisedagain{The} user-defined nature of functions in the serverless paradigm implies that the context-specific solutions have limited applicability. Moreover, the high-level abstractions offered by the serverless paradigm imply that the resource allocation and its related optimizations are accomplished by the platform, whereas, approximate computing techniques often require direct access to the hardware resources. As such, while users in the HPC, Grid, and IaaS systems can tweak their tasks to exploit the approximator hardware~\cite{sim2018dps} or use \revisedagain{Dynamic Voltage Frequency Scaling (DVFS)} techniques \cite{rahimi2015approximate} \revisedagain{with some control over the underlying hardware}, the serverless paradigm abstracts these aspects from the user\finalrevise{. Hence,} \revisedagain{if such capability is to be offered, it'd be} the serverless platform’s responsibility to provide them as \finalrevise{an} \revisedagain{abstracted} platform feature.

\begin{figure}[!ht]
\centering
    \includegraphics[width=0.55\textwidth]{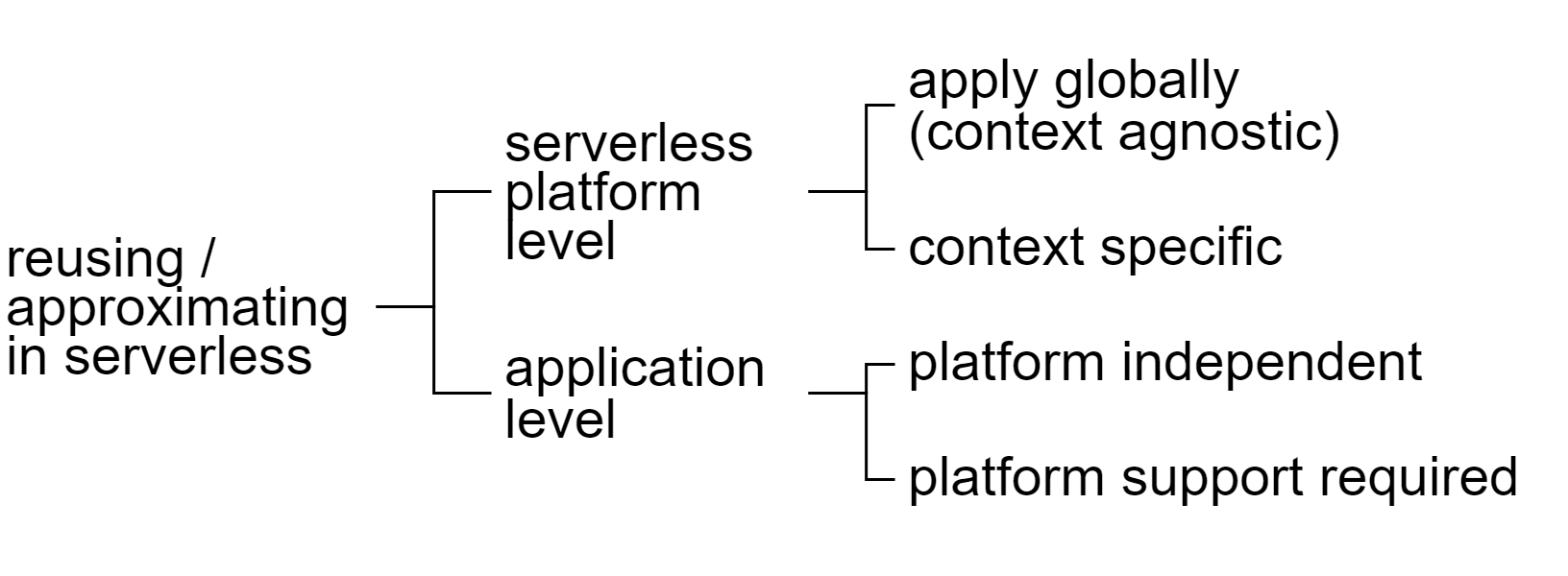}
        \caption{\small{Categorizing reusing and approximation techniques that can be used in the serverless paradigm.}}
                \label{fig:Uniqueness}
\end{figure}


\revisedagain{\finalrevise{Generality and supporting user-defined task-type are the main selling points of the serverless paradigm}. However, these features also makes it challenging to apply reusing and approximation techniques on a serverless system with a wide variety of task types. Such techniques, in theory,} can be applied either from the serverless platform side, from the function (application) side, or collaboratively from both sides. As shown in Figure~\ref{fig:Uniqueness}, we can categorize the reusing and approximation techniques of the serverless systems into four types: (1) Global techniques applied at the platform level, regardless of the application context. For instance, container or container image reusing techniques \cite{zhao2020duphunter,rastogi2017cimplifier,vassiliadis2022fast} that can universally improve the performance, regardless of the task types; (2) Selective techniques applied to certain tasks (contexts) at the platform level. For instance, consider a result caching system \cite{SecureDedup} that reuses the results of recently executed tasks. Such techniques are only safe to apply to functions that are truly stateless, such that the function execution does not have side effects on the internal/external states; (3) Application-specific techniques independent of the serverless platform. For instance, consider an application that can approximate the output precision according to the input and context, regardless of the underlying platform; and (4) Application-specific techniques with platform support. For instance, consider a function that can do approximate computing via ASICs. The function must be developed to allow such approximation, and the serverless platform must recognize when to assign such function to the ASICs.

In this survey study, we not only target the first two types of techniques that directly involve the serverless computing platform, but we also discuss the fourth type, which is application-specific techniques that require support from the serverless framework.
\section{Reusing Opportunities in the Serverless Clouds}
\label{sec:reuse}

\begin{figure}[!ht]
\centering
    \includegraphics[width=0.8\textwidth]{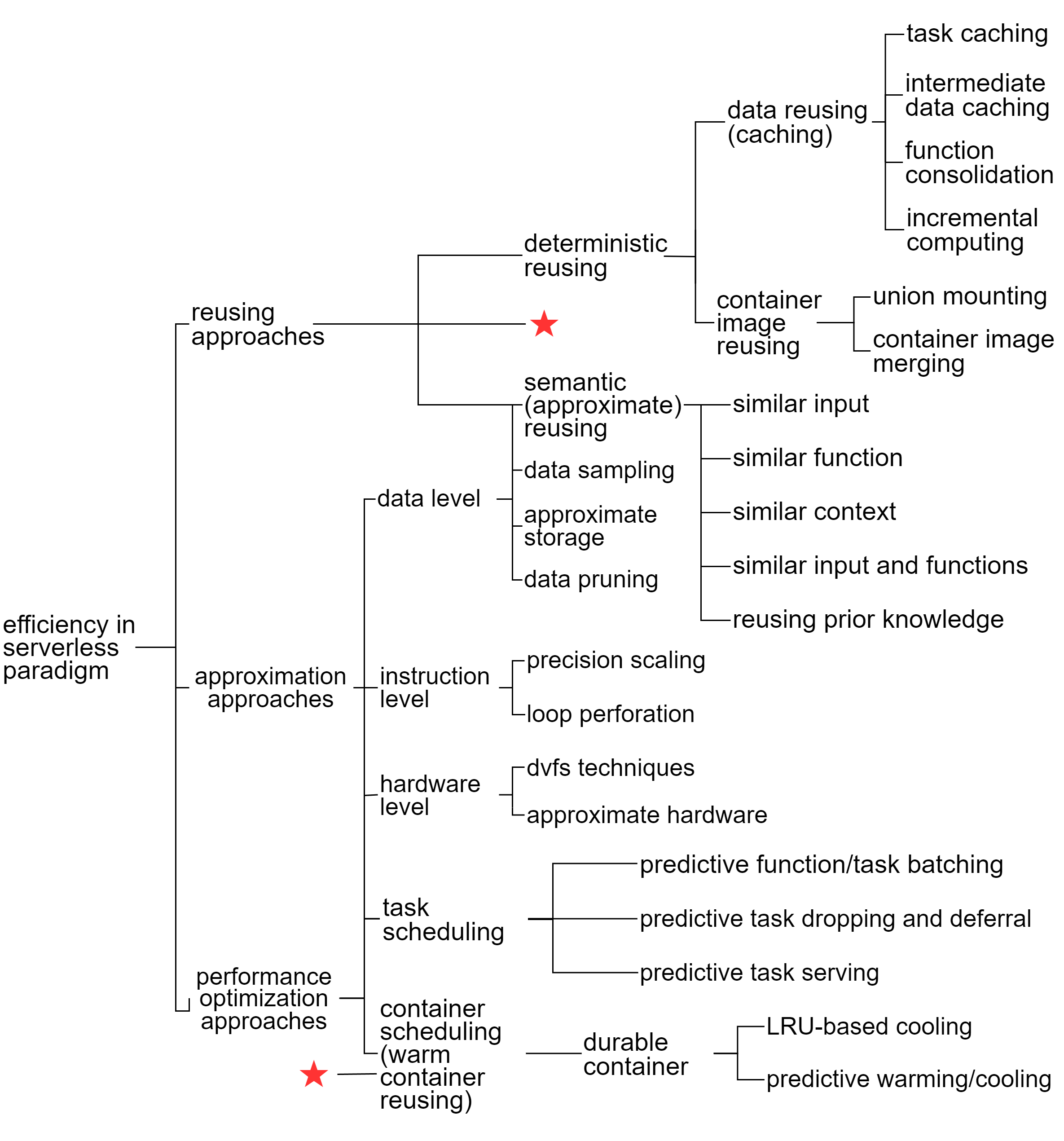}
        \caption{\small{Taxonomy of approaches for efficiency in serverless computing platforms. The approaches are classified into two categories that are based on the \emph{computational reuse}, and then based on \emph{approximate computing}. There are approaches in the intersection of these two, known as \emph{approximate reusing}.}}
                \label{fig:ReuseApprox}
\end{figure}

Reusing is defined as a way(s) to reduce resource usage and increase efficiency via deduplicating data or computations that share a certain level of similarity.
Historically, reusing (\eg in form of caching) has been a fundamental approach to achieving software and hardware efficiency. In this section, we study reusing in the context of serverless computing and describe how it can be potentially advantageous for both cloud providers and users. Then, we explore a wide variety of techniques to carry out computational reuse in the serverless context. A summary of these techniques is shown in Figure~\ref{fig:ReuseApprox}.

\subsection{Deterministic Versus Semantic Reusing}
\textit{Deterministic reusing} refers to the set of techniques that can detect reusable computation or data in a definitive manner and perform the reusing without altering the results of the involved tasks. That is, these techniques do not require inferring the semantic similarity between two data or two computations. 
An example of deterministic reusing is when a user requests for the re-execution of a stateless video encoding function with identical specifications on the same video. As the stateless function output depends entirely on its input, a re-execution of a task executed with the same input arguments yields the same result. Thus, the video result of such task execution can be cached and reused. The majority of the deterministic reusing techniques operate based on detecting frequently-used computation and/or data, then caching them to be reused at a later time~\cite{denninnart2020cost}. \revisedagain{The cached data can be stored in different locations, such as platform-level storage \cite{zhang2023infinistore}, node-level storage \cite{sreekanti2020cloudburst}, or inside the container in case durable containers are utilized (as described in Section~\ref{sec:durableFn}).} Accordingly, deterministic reusing techniques can be categorized into \emph{data reusing} (a.k.a. caching) and \emph{process reusing} that are elaborated in the next subsections.

Alternatively, \textit{semantic reusing} aims to find a semantic relationship between similar (\ie non-identical) data~\cite{krishnan2016incapprox} and perform reusing on them. In a sense, semantic reusing can be viewed as an approximation approach, due to similarity detection and the fact that the execution results of the involved tasks are not deterministic. However, the semantic similarity detection is uncertain and prone to misinterpretation, thus, can potentially lead to incorrect results. 

An example of semantic reusing can be in an application that provides ambient perception for blind and visually impaired users by detecting obstacles in their environment (\eg \cite{mokhtari22,satya14}). These users often visit repeated locations and interact with objects they have previously encountered during their day-to-day activities. The captured pictures of these objects are \emph{digitally} different because they are captured from different angles or under different light conditions. However, these pictures are \emph{semantically} representing the same objects \cite{guo2018potluck,guo2018foggycache}. Accordingly, a reusing mechanism that can pre-process incoming images and detect whether a semantically similar one has been (or is being) processed can be helpful. 

Further details about different approaches to performing semantic (approximate) reusing are discussed in Section~\ref{subsec:semreuse}

\subsection{Data Reusing}\label{subsec:reuse_result}
\emph{Data reusing} is defined as the act of saving certain data to be reused at a later time. It is an integral part of different levels of modern computing systems---from the hardware level to the compiler and execution levels. In the particular case of serverless computing and FaaS, the fact that tasks are fine-grain (function level) and stateless provides an ideal opportunity for data reusing via saving (caching) and using the results of function execution again.

Data reusing is predominantly achieved via caching operation. Caching is an optional, but very popular, operation in the computing systems to mitigate the slowdown resulting from accessing the storage systems, hence, accelerating the task execution. That is, the system can still function correctly, even if it misses the cached data and retrieves it from the storage. Since caching is limited and often costly, establishing a trade-off between cost and efficiency in the caching scheme is of paramount importance. An extensive cache space imposes a significant cost, whereas, an inadequate cache space leads to missing reusable function results that, in turn, increase the re-computation cost and the response time~\cite{jiang2017secure,arteaga2016cloudcache}. 

Based on the way data is stored and reused, the data reusing techniques fall into the following four categories that are elaborated in the next parts: (1) Task caching; (2) Intermediate data caching; (3) Function consolidation; and (4) Incremental computing solution. 


\paragraph*{Task caching}
Task caching is the act of capturing and reusing the result of a task (function) execution. \revisedagain{For stateless functions}, this caching technique can be deployed transparently from the user's perspective. The cached data can be quickly identified by making use of the hash value of the function call signature that is composed of the function name and its arguments \cite{mao2012cache}. The cache table can be either shared across users (\ie public) or maintained separately for each user (\ie private). Unarguably, a public cache table maximizes the likelihood of data reusability across functions of all users. However, it can be vulnerable to cache poisoning attacks \cite{wallenta2010detecting}. Alternatively, a private cache table offers better security via segregating the cache table either based on the user or the function~\cite{SecureDedup}.

\paragraph*{Intermediate data caching}
Intermediate data caching maintains partial results of the execution, rather than the final result. The technique is usually suited to  \revisedagain{a workflow} with multiple computing steps, in which caching the intermediate result offers a higher chance of reusability than caching the final result~\cite{arteaga2016cloudcache,wu2020autoscaling,sreekanti2020cloudburst}. \revisedagain{For example, consider a two-stage calculation where the result of the first stage is fed to the second stage. If the first stage only has a few possible input parameters that take a long time to compute, and the second stage is fast, but has a large domain of possible input parameters, then, caching the result of the second stage yields a poor cache-hit, whereas, caching the result of the first stage (intermediate result) leads to a more reusable caching. In a serverless platform, } intermediate result caching can be carried out via a key-value storage for the (stateful) functions \cite{carlson2013redis}. To achieve reusability, this caching technique requires the function code to explicitly store and retrieve partial results from the caching system. Thus, it is not transparent from the user's perspective.

\paragraph*{Function consolidation}
Another variation of data reusing occurs in the function workflows~\cite{wang2020accelerating,popa2009dryadinc,casas2017balanced} that includes multiple functions with data dependencies between them. In such workflows, because each function can be potentially allocated on a different machine, the overhead of transferring the output of one function to be used as the input of another function can be significant \cite{wang2020accelerating}.  Such overhead can be mitigated by fusing functions. That is, two or more functions can be consolidated to form one function, such that the whole function is executed together and the data transfer overhead is eliminated.  Let \texttt{A+B} represent a consolidated version of two functions \texttt{A} and \texttt{B}. In \texttt{A+B}, the output data of \texttt{A} can be reused by being directly fed as the input of \texttt{B}. We note that function consolidation has another benefit of scheduling one task as opposed to two individual tasks, hence, in Figure \ref{fig:ReuseApprox}, it can be also considered under the ``task scheduling'' category. However, given the increasing popularity of function workflows in modern software engineering paradigms and the substantial overhead of data transfer across data centers, we believe that it better fits the ``data reusing'' category. 

Function consolidation can be employed at the programming level via defining less granular functions, however, doing so is against the microservice-based software engineering methodology and makes the function maintainability cumbersome.
A more efficient approach for function consolidation is to let users maintain their fine-grain functions and let a framework in the back-end automatically carry out the fusing~\cite{akkus2018sand,dukic2020photons,wang2020accelerating} process without any user intervention. The main challenge in this approach is how to balance reducing the data transfer overhead against the resource inefficiency potentially caused by forming coarse-grain (consolidated) functions \cite{wang2020accelerating}. The reason for inefficiency (and potentially resource wastage) is that coarse-grain functions limit the ways tasks can be allocated, thereby, reducing the flexibility of resource allocation methods. For instance, consider a workflow with two chained functions where the first one, denoted $f_m()$, is memory-intensive and the second one denoted $f_c()$, is CPU-intensive. These functions can be efficiently allocated on different machines in a heterogeneous system---$f_m()$ on a memory-rich and $f_c()$ on a CPU-rich machine. However, consolidating the two functions requires a machine that is both memory- and CPU-rich. This can potentially lead to inefficient resource utilization or the incurred cost of accessing such a high-end machine can be more than the benefit of reducing the data transfer overhead. In summary, we note that an efficient function consolidation framework must consider the function's characteristics and bases its decisions on comprehensive function profiling \cite{shahrad2020serverless}.


\paragraph*{Incremental computing}
The fourth category of data reusing is the incremental computing technique~\cite{krishnan2016incapprox}. Similar to task caching, this technique also caches the task result. However, the cached content is reusable beyond the tasks with identical input arguments. Incremental computing utilizes a \emph{correction function} to adapt (\ie prune and expand) the cached results based on the new input. A common use case of incremental computing is in data analytics~\cite{tiwari2014mapreuse,krishnan2016incapprox,hersche2022constrained}. For instance, consider a repetitive function (query) that is regularly applied against a database with minor daily \revisedagain{ changes (\eg average number of active users in the past 365 days)}. A na\"\i ve way to handle the query is to thoroughly search the database every time. Alternatively, an incremental reusing technique retrieves the results of the prior \revisedagain{period (\eg 365-days result calculated yesterday)} and corrects them by pruning the invalid records \revisedagain{(\eg data from 366 days ago)} and adding new results \revisedagain{(\eg data from today)} via searching only within the updates in the database since the prior \revisedagain{period.}  It is noteworthy that incremental reusing is a highly context-specific technique, and currently, it is not implemented within the general-purpose serverless platforms. For instance, Zhang \etal~\cite{zhang2021serverless} propose a serverless and FaaS-based platform that takes advantage of incremental computing in the video analytics context. Their use case employs a deep neural network (DNN) model for video object classification. The model requires frequent updates to its weights to gain maximum accuracy with the ever-changing datasets. To avoid the excessive cost of re-training the model frequently, their platform deploys the incremental machine learning~\cite{wu2019large} technique to keep up with the gradual changes in the input datasets.

One way to enable incremental computing in future serverless platforms is to allow users to define multiple auxiliary functions, in addition to the main function. The auxiliary functions can include: a \emph{subtract function} (to remove part of the existing results that are not valid for the input argument) and an \emph{addition function} (to include new results to the existing ones and adapt them based on the new function arguments).

\subsection{Container Reusing}
\label{subsec:reuse_process}
Apart from the data reusing, serverless efficiency can be gained via reusing at the sandboxing platform (\ie container) level. More specifically, container reusing can be carried out at the \emph{container image} level that is described in the following subsection or at the \emph{container instance} level that is described in Section~\ref{subsec:contreuse}.

\subsubsection{Container Image Reusing}
\paragraph*{Union mounting}  
Union mounting \cite{huang2019fastbuild} is a well-established approach to reduce the container image footprint and, subsequently, the container start-up overhead. In union mounting, a container image is stored as multiple separate layers that can mount together to form an image. 
In this manner, the same layer can be utilized, \ie reused, in multiple images that share a module, thus, the storage overhead of container images is reduced. For instance, two machine learning functions can share the same operating system layer and the same machine learning framework (\eg TensorFlow \cite{abadi2016tensorflow}) layer. Then, they only differ on the application and model layers. Although deduplicating redundant layers is already widely used by the runtimes, further research works have recently been undertaken to improve the efficiency of deduplicating~\cite{zhao2020duphunter,rastogi2017cimplifier} and to extend the deduplication idea to reuse \emph{similar} layers~\cite{saurabh2019semantics}. Notably, Zhao~\etal's Duphunter~\cite{zhao2020duphunter} is a replacement layer loading module for the Docker platform. The architecture is more effective in deduplicating similar layers across multiple docker images with less deduplication overhead than prior designs.

\paragraph*{Container image merging}  
When two or more functions whose container images have only minor differences are launched from a cold start, union mounting can capture and reuse several parts of their images. However, from the serverless platform perspective, these two are separate functions, hence, are treated independently. If these functions are infrequently invoked, they can get evicted from the memory. 
To encourage the system to keep these infrequently-used functions in the memory, a.k.a. \emph{warm}, one approach is to merge these functions such that they share the same container image. In this case, the collective invocation frequency of these functions is increased that can avoid the memory eviction for them.

\subsection{Semantic (Approximate) Reusing}\label{subsec:semreuse}
While deterministic reusing is only built on the identical data, semantic reusing aims to achieve reusing where the data are not digitally identical, but the base for reusing is some form of \emph{semantic similarity} between the data. Semantic reusing can maintain semantic correctness by producing approximately the same results, while remarkably avoiding resource wastage and improving users' satisfaction. That is why this category of reusing is also considered a type of approximation that is discussed in  Section~\ref{sec:approxcloud}. However, the mechanism to detect semantic similarity is not infallible and, similar to other approximation approaches, can potentially lead to inaccurate results that in certain contexts (\eg video streaming) could be tolerable and still useful. The semantic similarity detection and reusing functions are often application-specific. However, similarity detection is more efficiently performed at the scheduler level rather than the application (container) level. To enable wide adoption of semantic reusing, one potential way in the future serverless platforms is to provide a standardized API for users to define the semantic similarity detection for their function to be utilized by the platform scheduler.  

In the serverless context, there are four types of similarities that can be leveraged for semantic reusing:
    \paragraph*{(1) Similar input:} For stateless functions, the result of execution depends only on the input argument. The input is often composed of multiple parameters. For instance, a video encoding function has the video segment, resolution, frame rate, bit rate, and codec as its input parameters. By designating certain parameters of the function as approximable, function reusability can be enhanced~\cite{guo2018potluck}. For instance, consider two users who call the \texttt{transcode} function to process the same video with two different (but approximable) resolutions. In this case, the system can approximate the resolution arguments and process the function once, instead of twice, and send the results to both users. Such scenarios can be particularly useful under certain circumstances, such as when the system is oversubscribed or when some users can tolerate lower QoS, such as free subscribers of a video streaming service.

    \paragraph*{(2) Similar function:} Following the microservice-based software development principles \cite{arachchi2018continuous}, generally, functions are developed to be short and single-purpose to ease the continuous deployment (CD) process. Therefore, it is likely that multiple users define similar functions that try to achieve the same purpose. These functions are semantically the same while having distinct source codes. Let functions $A$ and $B$ be semantically the same and $x$ be an arbitrary input argument. In this case, the serverless computing system with the function similarity detection mechanism in place can reuse the result of $A(x)$ for $B(x)$ too. Moreover, since these functions are similar, one of the functions can be replaced by the other one, rather than keeping both functions available. This saves the number of functions that have to be kept active for rapid execution.
    
    \paragraph*{(3) Similar context:} In this category, the state, a.k.a. context, of a stateful function is to be reused. This is particularly helpful in the circumstances that the state data is not sensitive and can tolerate minor differences. That is, minor changes in the state do not significantly impact the results. For instance, consider a function for online learning of an image classification machine learning (ML) model \cite{redmon2018yolov3} where the state data is the weights of the ML model. Other functions that intend to train the same model can reuse the state (weights) of the model from the earlier function, and train it further with their data. Such reusability makes the ML model converge faster and is more cost- and energy-efficient. A similar type of reusing can be considered in a serverless federated learning system \cite{qu2021serverless} where the workers reuse a central model and train it further with their local data.
    
    
    \paragraph*{(4) Reusing prior knowledge:} In a serverless system, function (task) profiling data, collected and summarized from prior executions via automatic task profilers \cite{seneviratne2011task}, can be supplied to the task scheduling module to maximize the resource allocation efficiency of the system \cite{spe22bader}. Moreover, for a new user-defined function that lacks prior profiling information, the serverless platform can reuse prior knowledge of semantically similar functions to estimate the execution time of the new function on different machine types available in the system. Otherwise, lack of such information causes uninformed scheduling decisions that negatively impact the users' perceived QoS. Unlike other forms of semantic reusing that directly impact the quality of the results, reusing prior knowledge deals with the system parameters, such as utilization and QoS, and is used to improve them.

    Transfer learning is a technique to reuse the knowledge gained while solving one problem and applying it to a different but related problem \cite{tan2018survey,li2022perspective}. Accordingly, transfer learning can be employed to predict the execution time of a new function on a given machine type based on the trained networks of other functions on the same machine type. Moreover, methods can be explored to measure the semantic similarity between the new function and each one of the existing functions based on the dependencies and libraries shared between them. Then, the weighted average similarity of the new function with other functions can be used to estimate the execution time of the new function on that particular machine type. It is worth noting that, once enough execution time information for the new function is collected, a model specific to that task type can be trained to infer its execution time independent of other task types. Similarly, when a new machine type is added to a heterogeneous serverless system, the prior profiling information of functions on other machine types can be leveraged to estimate the expected execution time of the functions on the new machine type, thereby, utilizing it more efficiently.


\section{Approximate Computing in the Serverless Clouds}
\label{sec:approxcloud}
Approximate computing allows functions (tasks) with unaffordable response time, energy, or cost constraint(s) to be completed within its constraint(s)~\cite{jayakumar2016energy}. Even the tasks with affordable constraints that can tolerate approximate results, a.k.a. multi-fidelity tasks~\cite{satyanarayanan2001multi}, can use approximate computing to bring about further resource-saving in the system. Since approximate computing compromises the precision and/or accuracy of the results, we envision that in comparison to reusing, employing approximation has less scope in the serverless platforms and is only to meet the tasks' constraints that are otherwise unattainable. Some notable use cases of approximate computing in the serverless systems include:
\begin{itemize}
    \item[1)] Improving the \emph{response time} via fast approximate results, before confirming or correcting the results by the exact computing.
    \item[2)] Providing an approximate result to reduce the incurred \emph{cost}.
    \item[3)] Providing an approximate result only if the system is \emph{oversubscribed} and cannot perform exact computing in time.
\end{itemize}

There are various approaches for approximate computing that can improve the efficiency of serverless computing platforms. We can categorize these approaches into four classes, as shown in the lower part of Figure~\ref{fig:ReuseApprox}. In this section, we first position approximation approaches in comparison to the reusing-based approaches and then, discuss the general requirements for function approximation in the serverless context. Next, we elaborate on the four categories of approximation in  Sections~\ref{subsec:datalevel}---\ref{subsec:schedlevel}.
\subsection{Approximation Versus Reusing}
The main difference between approximation and computational/data reusing is the impact on the result accuracy and precision. Computational reuse accelerates the turnaround time or saves the computing resource without influencing the task result. Conversely, approximate computing compromises the result accuracy and/or precision in favor of time- and/or resource-saving. 

While approximate computing can be applied to tasks independently, many of the approximation techniques benefit from reusing information gathered from prior tasks. Such information can be either predefined ahead of time, \eg trained ML models, or collected and applied dynamically at the run time, \eg caching the results of similar tasks. Approximate computing techniques that directly reuse the result of other similar tasks are also known as approximate reusing which is discussed in the previous section.

\subsection{Approximate Computing Requirements}
Approximate computing exploits the \textit{resilient} property of the system by producing inexact but acceptable results at a lower cost. A resilient system~\cite{jayakumar2016energy} or application should be able to tolerate a certain amount of deviation from the ideal result~\cite{chippa2013analysis}. Specifically, an approximate computing technique should not cause deviations that exceed the application resiliency in both error magnitude and likelihood.

\paragraph{Error magnitude}
Error magnitude is defined and measured based on the variation of the obtained result from the ideal result. Applications' tolerance to the error magnitude varies based on the context. For example, video processing for live video conferences can tolerate a higher error magnitude than the video processing for traffic cameras that has to perform vehicle identification. 

\paragraph{Error chance}
Another dimension of error quantification is the likelihood of error occurrence. Formally, the likelihood of getting an overly inaccurate approximation is called the \emph{error chance}. Certain approximation techniques (\eg DVFS \cite{rahimi2015approximate}) produce mostly accurate results, however, there is a chance that the error occurred, and the result accuracy is off by far beyond the acceptable error magnitude. In such approximation techniques, upon detecting an error by a validation function, a correction function \cite{hepworth2014model} is employed to fix the error retroactively. If the chance of getting an error is considerably high, then the overhead of correcting the results frequently can exceed the benefits of the approximation.


\subsection{Data-Level Approximation Approaches in Serverless Computing}\label{subsec:datalevel}



    
 
\subsubsection{Approximate Reusing}
Approximate reusing is the same as semantic reusing and is performed via identifying potentially reusable tasks and using their data to approximate other tasks~\cite{samadi2014paraprox}. Allowing repeated function calls to reuse the result of similar, but not strictly the same, tasks promotes reusability. The main challenge in approximate reusing is detecting the semantic task similarity. Applying this type of computational reuse can improve both the user and system metrics unless the users opt-out of it due to privacy concerns.
The feasibility of this approach entirely depends on the feasibility of the semantic similarity detection system. As such, they are mostly applicable in domain-specific serverless systems, for example in video streaming \cite{denninnart2022smse}. 


\subsubsection{Data Sampling}
For functions that work on a batch of data, such as data analysis works \cite{pu2019shuffling,krishnan2016incapprox}, it is possible to reduce the input data size by sampling from the dataset. Various techniques have been explored for data sampling, performing computation on the samples, and then analyzing the variability and the error rate in comparison to the complete data analysis on the whole data. For example in the IoT context, ApproxIoT~\cite{wen2018approxiot} provides a method to sample data from a stream of unknown data sizes. Sampled data are stored in a size-limited reservoir. New data can randomly replace the existing ones in the reservoir. We believe this approach of approximate computing can be offered as an optional service for various special-purpose stateful serverless functions.

\subsubsection{Approximate Data Storage and Data Pruning}
Approximate data storage can be achieved via either persisting only a portion of the data or scrambling multiple data points together. For instance, in a serverless multimedia cloud \cite{li2018cost,denninnart2021harnessing} only base video formats can be persisted, and other less popular formats can be transcoded lazily---upon receiving a user request. 
For the content types that are error sensitive, similar data points can be stored together, \ie merged, via lossless data compression methods \cite{gupta2017modern}, whereas, for the content types that can tolerate minor errors, \eg multimedia and image, lossy compression methods \cite{zheng2018error} can be employed to approximate similar data, thereby, carry out a more effective compression.

Instead of involving the user in handling data storage and reusing, the serverless platform can be made aware of the services that store and deduplicate similar data~\cite{paulo2014distributed}. By performing the data management at the platform level, the platform can maximize the likelihood of detecting similar data and performing deduplication. In such a storage mechanism, the process of retrieving data can also be approximated to reduce the data retrieval overhead. Eventual consistency~\cite{balegas2015putting} can be utilized on the data that is accessed by multiple tasks. Such a relaxed consistency control incurs a lower cost than a strict data consistency at the price of allowing the task to start with inconsistent data. Serverless functions are generally short-lived and are easily undoable. Therefore, tasks that start with inconsistent data can be canceled and restarted with minimal overhead.

\subsection{Instruction-Level Approximation in Serverless Computing}
While instruction-level approximation techniques seem to be very context-specific and are applied at the application level, they can benefit from the metadata provided by the serverless platform. For example, the serverless platform can leverage resource utilization information to determine the normal or approximate processing of a given function.

\subsubsection{Precision Scaling and Stochastic Computing}
The earliest forms of approximate computing were built by necessity in computer storage systems that could not store infinite decimal points \cite{pokhilko2018double}. Hence, the numbers had to be approximately stored and represented by a close-enough value. 
Then, the concept was further developed to a more deliberate dynamic precision scaling~\cite{zamani2017edge} where calculating the precision is scaled based on multiple factors, including the trade-off between computing precision and energy or turnaround time requirement.

Stochastic computing \cite{salehi2016stochastic} is a popular collection of techniques to achieve precision scaling via representing a continuous value in form of a stream of bits. In this case, calculating the precision can be scaled by altering the number of bits in the bit stream. Making use of stochastic computing often implies designing a domain-specific processing unit (a.k.a. ASICs) hardware.
Since precision scaling requires a specific framework and stochastic computing requires specialized hardware, these techniques have not yet gained wide adoption in the cloud computing industry. However, with the increasing prevalence of ASICs and, particularly, the trend in using precision scaling for the ML inference on the energy-limited edge devices \cite{vogel2017efficient}, we envisage that these solutions will eventually carry over to the cloud systems too. As such, serverless solutions to support domain-specific processors that achieve instruction-level approximation will be demanded shortly to hide the complexity of deploying ASICs across the edge-cloud continuum from the user perspective. Such solutions will help the users to become solution-oriented and focus on their application logic, rather than dealing with the operational details of different hardware systems.

    
 


\subsubsection{Loop Perforation and Instruction Replacement}
In a serverless computing platform, functions can either be provisioned as containers or, more popularly, as functional code blocks. The serverless platform can analyze the user code and  optimizes them to save the computing resources. We believe that approximating frameworks such as Approxilyzer~\cite{venkatagiri2016approxilyzer}, proposed by Venkatagiri~\etal can be deployed as an optional service to achieve such optimizations in serverless platforms. Approxilyzer analyzes the machine code of the function and dynamically replaces parts of the instruction with the approximated version. The aggressiveness of the approximation can be tuned based on the demanded quality, resiliency, and the approximation overhead. 

\subsection{Hardware-Level Approximation In Serverless Computing}
When heterogeneous computing is supported in a serverless computing system, specialized hardware that allows approximate computing at the hardware level can be offered as one of the resource types. The offering can be especially attractive for use cases, such as big data and machine learning, that are data-intensive and can benefit from domain-specific machines to offer low latency and real-time services~\cite{du2014leveraging}. Moreover, making use of specialized hardware to accomplish approximate computing can effectively reduce the energy consumption and footprint of cloud datacenters~\cite{du2014leveraging}.

Two main hardware-level approximation approaches are namely, Dynamic Voltage Frequency Scaling (DVFS)~\cite{rahimi2015approximate} and approximator hardware. DVFS is a technique that strategically under-volt common hardware systems. Although such Undervoltage induces errors in the computation, applying it in a controlled manner, such that the error rate is tolerable by the applications, can improve the efficiency of the serverless system. For instance, Rahimi \etal~\cite{rahimi2015approximate} propose to strategically under-volt the GPU in favor of energy efficiency, while employing Hamming distance \cite{bookstein2002generalized} to allow more error tolerance at the application level. 
On the approximator hardware side, certain computations are common and can greatly benefit from the approximator hardware. A few notable examples of such tasks include stochastic computing that can greatly accelerate computing by its approximator hardware~\cite{sim2018dps}, as explained in the previous section; DNN approximate inference that uses specially designed inference hardware~\cite{vogel2017efficient}; and image encoding using the approximate encoder hardware~\cite{snigdha2016optimal}.

The list of tasks that has approximator hardware support is still expanding, as more use cases are found to benefit from the approximator hardware and more tools to aid approximator hardware emerges~\cite{yazdanbakhsh2015axilog}. Accordingly, we envision that the approximator hardware will find its way to the serverless systems in the future and their platform should be able to accommodate them transparently and efficiently.

\subsection{Scheduling-Level Approximation in Serverless Computing}\label{subsec:schedlevel}

\subsubsection{Task Scheduling Approaches}

\paragraph*{Predictive function (task) batching}
Although the request turnaround time is one of the main criteria in measuring the performance of serverless clouds, not every application needs the function to complete as soon as possible. Moreover, even deadline-constraint tasks often can tolerate some delay, a.k.a. slack, before missing their deadlines. 
A recent study conducted by Eismann \etal~\cite{eismann2020serverless} demonstrates that around 38\% of their surveyed serverless applications have no latency requirement and another 28\% of the applications have a few latency-sensitive functions. Only 2\% of the applications are real-time with rigid latency constraints. 

To maximize the efficiency of serverless systems, the user or the system should have a way to declare the task urgency, possibly in multiple tiers (\eg urgent vs non-urgent) or as a continuous number, such as a deadline value. In this case, highly urgent tasks can be scheduled to complete with the minimum turnaround time via warmed containers or by some form of approximation, whereas, the less urgent ones can potentially wait to aggregate with other similar arriving requests, thereby, maximizing the container reuse and reducing the incurred cost. 

The scheduler must have the ability to predict the cost-benefit of delaying tasks in favor of batching them, such that each task waits as long as possible without missing its deadline to share the function container and other resources with other tasks. Predictive function batching is a viable technique that can be implemented in the existing serverless systems and multiple research works, \revisedagain{ though not directly targeting serverless cloud, can be applied to serverless as well. For instance,} Grandslam~\cite{kannan2019grandslam} scheduler dynamically reorders tasks to maximize the task batching and minimize Service Level Objective violations in an oversubscribed system.
Fifer~\cite{gunasekaran2020fifer} includes a scheduler with mechanisms to batch tasks and reduce the amount of container usage and cold start overhead within a given latency budget. Unlike Grandslam, Fifer tries to minimize the resource usage in an underutilized system, rather than trying to meet the tasks' deadlines in an oversubscribed system. \revisedagain{The core idea of these two works can be applied to the serverless engine's task scheduler.}

\paragraph*{Task dropping and deferral}
In a serverless system, each task request can be part of a bigger workflow. In some use cases, the workflow includes optional steps (tasks) whose loss can be tolerated~\cite {yin2018data}. Such a feature can be exploited at the scheduling level, particularly, to mitigate resource oversubscription \cite{gentry2019robust} via dropping the optional tasks~\cite{mokhtari2020autonomous} or deferring their execution to a later time when the system is less busy~\cite{gentry2019robust,denninnart2019improving}. One use case that can take advantage of such workflow level approximation is in video conferencing where the voice quality enhancement task on the received video segments can be skipped, \ie dropped, to keep up with the liveness of the streamed video contents \cite{li2018cost}. While this technique is feasible, it is currently not widely adopted on cloud platforms because: (a) dropping tasks intentionally can impact SLO compliance; and (b) on the cloud, it is commercially viable to scale out and cover the surge in demand, rather than dropping tasks to fit the resources. However, in an emergency, such as disaster recovery, or the resource- and energy-limited edge systems, task dropping and deferral techniques can be instrumental to improve the overall performance of the system. \finalrevise{ Due to the nature of user-defined and time-sensitive task types (with deadline), we believe that such desperate techniques are not yet practical to be offered on a general-purpose serverless framework.}

\paragraph*{Predictive task serving}
The scheduler of a serverless computing platform can operate proactively and approximately anticipate the arrival of the user requests.
\revisedagain{For instance, Roy~\etal~\cite{roy2022daydream} proposed an approach to predict the function calls via fitting the invocation trend to a statistical distribution.}
Such prediction can help the platform to pre-warm function containers and load their required data in speculation of upcoming task requests. 
 Predictive task serving is done internally on the large-scale cloud providers~\cite{shahrad2020serverless}. However, the exact details on how they carry it out are often not publicly available.

\subsubsection{Container Scheduling Approaches}

\paragraph*{Durable container}
\label{sec:durableFn}
Prior studies~\cite{eismann2020review,shahrad2020serverless} express that, in a serverless system, a certain percentage of functions are invoked frequently. If warm start execution of these functions incurs a significant loading overhead, then the frequent warm starts can compound a substantial overhead for the system. A common example of a function with high warm start overhead is the one that involves an online machine learning model as its state \cite{torrey2010transfer}. To solve such inefficiency, certain serverless computing platforms allow functions to run as a ``durable container'', a.k.a. non-transient container, which means the container does not terminate after the task completion. These durable containers can maintain the state across function calls (\eg updating  the machine learning model in the above example) that help to process subsequent tasks without paying repeated start-up overhead. In the event that there are multiple requests for the same function, the requests are queued to receive the service. Microsoft Durable Functions \cite{msAzureDurablefunctions} and Oracle Fn Flow \cite{fnflow} support this capability. We envision that the future serverless platforms will auto-detect frequently-used functions and make their containers permanent without any user intervention.

\paragraph*{Predictive warming and cooling}
\label{subsec:contreuse}

Due to the memory limitations of the servers, not all function containers can be maintained in the memory to rapidly start the functions' execution. Infrequently used functions have to be offloaded to the storage to make room for the frequently-used ones. Such a memory contention across function containers is one of the challenges in the serverless domain and resolving it entails dealing with multiple problems: (1) how to reduce the cold start time overhead? (2) how to keep more containers in a given memory space? (3) how to minimize the number of cold starts through efficient memory allocation?

The main approach to mitigate the cold start overhead is to alter the transient nature of containers and keep them in the memory even after the function execution. Another approach is to proactively load the container, \ie before the function invocation \cite{daw2020xanadu}. In this direction, research has been conducted by Shahrad~\etal~\cite{shahrad2020serverless} who studied 14 days of function invocation patterns in Azure cloud and leveraged that to develop a container memory allocation strategy. They propose to reduce the number of cold starts via categorizing the functions where each category has its pattern of \emph{pre-warm} and \emph{keep-alive} periods. Right after finishing an invocation, the function is removed from the memory for the \emph{pre-warm} period, because the system does not expect to get another invocation of the function shortly. Then, once the pre-warm period expires, the container is loaded back into the memory (\ie warmed) to get ready for the next warm start function invocation. If the function stays in the memory for more than the keep-alive period without any invocation, then the function is removed from the memory. Such a strategy reduces the number of cold starts for the majority of the functions, however, there are still some functions whose invocation pattern is unpredictable, hence, cannot benefit from the predictive memory allocation. Nevertheless, the benefit of correct predictions and delivering a warm start remarkably exceeds the cold start misprediction plus the solution overhead.

\section{Potential Future Research Directions}
\label{sec:app}
\begin{figure}[!ht]
    \centering
    \includegraphics[width=0.6\textwidth]{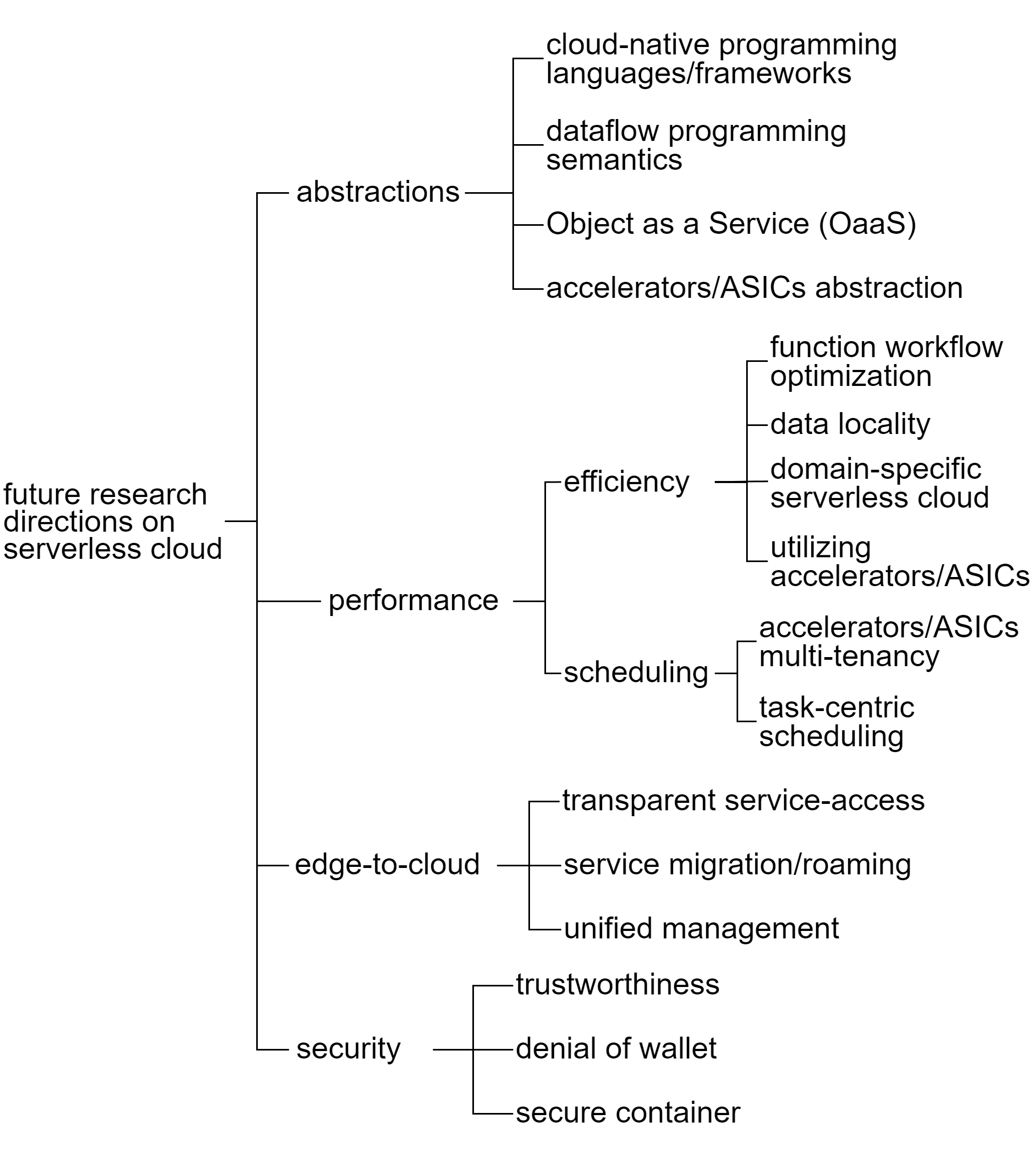}
    \caption{\small{Topics for future research directions on the serverless cloud.
    }}
    \label{fig:futurework}
\end{figure}

Serverless computing and FaaS are considered as the second generation of cloud computing systems. As such, it is crucial to know the directions that which this area of distributed computing is evolving. In this section, we discuss the research directions that have the potential for further exploration and are deemed as the enablers of the next-generation cloud computing systems. 
Figure \ref{fig:futurework} categorizes and summarizes these research directions into the four following thrusts: (A) providing higher-level abstractions to streamline and accelerate serverless application development; (B) improving the performance of serverless systems; (C) extending the serverless paradigm to encompass the edge-to-cloud continuum; and (D) improving the security aspects of the serverless systems.  

\subsection{High-level Abstractions for Serverless Computing}

\subsubsection{Cloud-Native Programming Language/Framework}

As the second generation of cloud technology is evolving based on the serverless paradigm and mitigating the programmers' job, we envisage that the technology is approaching the \emph{cloud-native programming languages} era. In such programming languages, FaaS concepts will be integrated into the programming languages and compilers can natively support them. Like that, functions can be defined as \texttt{cloudable} and seamlessly deployed on the cloud by the compiler or auto-migrates to the cloud by the run-time system to overcome the resource shortage or to achieve reusing and approximation. Moreover, these languages can integrate the ``programming code'' with the ``deployment code'', a.k.a. infrastructure in code \cite{tankov2021infrastructure}, that are currently developed separately by different people. In this manner, the program is analyzed to automatically identify the infrastructure demands and deploy them on the cloud, \revisedagain{such that the likelihood of reusing and approximation are maximized.} 

These features of the forthcoming cloud-native programming languages will democratize cloud programming, such that people without cloud knowledge can utilize them in their programs.

\subsubsection{Dataflow Programming Semantics}

 Although Function-as-a-Service (FaaS) abstraction relieves users from the burden of resource management (\eg load balancing and elasticity), it is not truly serverless, because it falls short on abstracting data management and the users still have to get involved with other services, such as AWS DynamoDB \cite{awsdynamodb} or AWS SAM \cite{awssam}, to serve desires of the functions and/or applications~\cite{pu2019shuffling,klimovic2018pocket,schleier2020faas}. In particular, in some use cases, the data can represent the function state. For instance, the state of an online-learned Machine Learning (ML) model is updated frequently. Embedding such a model into the function container image is not practical. Furthermore, ML models are large enough that cannot be fed as the function arguments, otherwise, the function startup overhead would become substantial.
 
In this scenario, the ML model is the function's state that is best to be stored in a synchronized storage service that can be accessed seamlessly upon the developers' demand. In this manner, the developers do not need to think about scattering, \revisedagain{reusing and} scalability of the state data on various storage or database services, instead, they can focus on the business logic of the application \revisedagain{while the system automatically manages data reusing and caching as it sees fits}. We believe that future serverless platforms will accommodate such dataflow semantics. 


\subsubsection{Object-as-a-Service (OaaS): Going Beyond the Function Abstraction}
Current serverless and FaaS solutions are not designed for (and cannot natively support) data-centric applications or \textit{state data}. The developers have to intervene and undergo the burden of managing the application data using separate cloud services, \eg AWS RDS \cite{awsrds}, to persist the state information.
Apart from the data aspect, current FaaS systems do not offer any built-in semantics to limit the access to the internal, a.k.a. private, mechanics of the functions. However, providing unrestricted access to the developer team has known side effects, such as function invocation in an unintended context, and data corruption via direct data manipulation. To overcome such side effects, developers again need to intervene and undergo the burden of configuring external services, such as AWS IAM \cite{awsiam} and API gateway \cite{awsgateway}, to enable access control.
This makes the development and maintenance of serverless applications difficult and cumbersome. To ease the management of data and accelerate the development of new services for them, a higher level of abstraction is desired that, in addition to hiding the resource allocation details, it can hide the details of access control and preserve its state from the users.

To natively support serverless data-centric application development, we envisage that in the future serverless platforms, the concept of \emph{object} can be borrowed from the Object-Oriented Programming, as the first-class citizen to encapsulate both computing (functions) and state (data) within a single object entity, and offer the notion of \emph{Object-as-a-Service} (OaaS) \cite{wis23}. OaaS will be the next generation of the BaaS part of the serverless system that not only will handle the state management and persistence with minor user intervention, but also will offer a high level of abstraction to the user. 

Not only do objects in OaaS offer encapsulation and abstraction benefits on top of the function abstraction, but they also unlock opportunities for built-in optimization features, such as data locality, data reliability, caching software reusability, and data access control. 
For instance, a serverless platform can offer the built-in encapsulation semantics for a cloud-based video streaming system \cite{denninnart2022smse}. In this system, video content is defined as an object that has the video file as its state and is bound to a set of functions that can be invoked by the viewer’s application that can potentially change the object state. A few examples of such services are as follows: Generating multilingual subtitles for the safety-related videos; Removing harmful and illicit content from the child-safe video content.

As mentioned in Section \ref{subsec:serverlessstate}, some stateful serverless solutions are implemented using the actor model. However, the major drawback of the actor model, considering no further optimization, is the limitation of reusing. OaaS can overcome this problem by leveraging the immutable data processing model. That is, upon invoking a function of an object, the OaaS platform outputs a new/updated object state, instead of updating the existing one. Implementing this semantic makes the function perform a stateless operation and enables the ability to apply approaches of reusing and approximation.

\subsubsection{Accelerators/ASICs Abstraction}
\label{subsec:accelabs}


    GPU computing gained a massive uptake as Nvidia continuously enhances GPU program-ability by abstracting hardware coding into user-friendly functions in CUDA. The abstraction makes it feasible to port CPU computing only code into GPU accelerated code. 
    Beyond GPU programming via GPU-specific code, is it possible to abstract serverless functions into hardware-agnostic code so that the framework can determine and utilize the appropriate hardware accelerator? For example, a large set of data can be processed one by one in the loop using mobile CPU, processed in batch using GPU, or approximately processed using approximator ASICs depending on what accelerator is available. Such capability can go a long way in popularizing heterogeneous computing in serverless clouds.

\subsection{Performance Aspects in Serverless}

\subsubsection{Memory Contention}

Unlike conventional cloud computing services (\eg IaaS services) where users are in charge of explicitly running and terminating the services, in serverless computing, the platform automatically allocates resources and runs the services upon request, and then de-allocates them when they are not needed anymore. This automated allocation and de-allocation of resources are realized via transient isolation platforms, \eg containers, that also enable charging users only for the times the services and functions are being used. Ideally, the containers should be \revisedagain{loaded from cold storage just once and then} maintained in memory to \revisedagain{guarantee} fast execution of the functions, which is critical for latency-sensitive applications \cite{baresi2017empowering}. However, in a large-scale serverless cloud, maintaining all the functions in memory is impractical, owing to both hardware and economic limitations. Hence, there is a memory contention between containers for memory access. Efficiently resolving this contention and determining the functions that should remain in memory \revisedagain{to maximize function reusing} is a challenging problem for the BaaS part of serverless systems.

\subsubsection{Function Workflow Optimization}

As mentioned in Section~\ref{subsec:trigger}, serverless functions can be in form of individual functions or as a workflow. Users can define a function workflow as multiple individual functions whose completion triggers the next function inline or define function workflow using a certain schema format. Currently, there are no platform-agnostic schema standards that we believe can be a useful development to maximize cross-platform compatibility, and streamlines cross-platform workflow migrations. 

From the scheduling aspect, big data analytics and map-reduce workflows are \revisedagain{observed} to perform poorly on serverless systems \cite{pu2019shuffling}. The reason is that the serverless task schedulers are generally not tuned for workflow tasks. \revisedagain{Functions with a large memory footprint are loaded to Perform a small task and then released, rather than being reused for other batches of data.} Serverless tasks are typically expected to have a short start-up time, hence, their schedulers are designed to be simple and lightweight. However, the next-generation serverless platforms are expected to consider particular scheduling arrangements for the function workflows. One approach can be scheduling the first tasks of the workflow upon arrival with the minimum startup overhead; for the following tasks, the scheduler  evaluates their data dependency with the other tasks and assigns them such that the data transfer overhead is minimized \revisedagain{and function reusing is maximized}.

\subsubsection{Data Locality Optimization}

Data locality optimization can reduce the wasted bandwidth and computing resources by minimizing the inter-rack data transfer to the minimum within the serverless cloud datacenter. \revisedagain{Such  optimization is critical for effective data reusing. In a poorly optimized system, \finalrevise{the} data transfer latency of the reusable data may \finalrevise{outweigh} the time required to recompute such data without reusing.} An example of such action is to place containers that frequently communicate within the same machine or rack, or schedule the tasks to be close to the data source.

While data locality is carried out by the user in the conventional cloud service models, it is the responsibility of the BaaS in the serverless paradigm. This is because the serverless aims at abstracting the user from the underlying resource management details and, as part of it, the user should not micromanage data locality. As the user grants full control of the process and data location to the cloud, the serverless platform has the luxury of optimally allocating tasks and data free from the user's constraints. For this reason, it is expected that the future serverless schedulers will leverage this flexibility to maximize data locality, thereby, reducing their cost and energy consumption.

\subsection{Efficiency via Domain-Specific Serverless Cloud Computing}
Along with the rise of domain-specific computing and ASIC hardware, domain-specific programming languages are also emerging for popular applications, such as machine learning, cryptography, multimedia processing, and even fluid dynamics \cite{friebel_fpga4hpc21}. Within this trend, we envision that the next step will be the emergence of domain-specific serverless cloud platforms for popular applications. A domain-specific serverless system will be equipped with specialized hardware (ASICs), support of special-purpose programming languages, and built-in domain-specific functions in its repository. Such platforms will expedite the application development process and shorten the CI/CD cycles. They will help users become solution-oriented and focus on their specific business logic, rather than spending time on developing basic services. In addition, they can support more flexible and application-specific billing schemes (\eg per successful transaction completion) with even cheaper prices, due to using more efficient specialized hardware. 

Importantly, a domain-specific serverless platform creates new scopes for serverless efficiency, via specialized function reusing and approximation techniques. For instance, a multimedia serverless cloud platform can harness its built-in knowledge of spatiotemporal multimedia streaming patterns to upraise the likelihood of function reusing.

\subsubsection{Utilizing Accelerator in Serverless Cloud}
Serverless computing systems were originally pioneered to create an illusion that there is no server to manage. Fulfilling this aim is viable when the underlying computing resources are either homogeneous or of the same architecture, a.k.a. consistently heterogeneous~\cite{gentry2019robust}. As such, to date, commercial serverless computing providers support only various forms of CPU-based machines that are auto-provisioned by the BaaS subsystem to match the function desires \cite{romero2021llama}. However, to accommodate more flexibility and efficiency in FaaS and serverless computing, inconsistently heterogeneous systems with a combination of CPU, GPU, TPU, FPGA, and other forms of emerging ASICs \cite{romero2021llama,nuclio} \revisedagain{(including hardware specifically \finalrevise{designed} for approximate computing)} are desired.  
With the slow-down in  general-purpose computing and the rise of specialized computing~\cite{thompson2021decline}, we expect that application-specific hardware becomes prevalent in serverless systems in the near future~\cite{romero2021llama}. 

Provided that the user functions can utilize heterogeneous resources through virtualization or heterogeneous-supported frameworks  (\eg TensorFlow for machine learning tasks, FFmpeg~\cite{2023ffmpeg,ffmpeg16} for multimedia processing tasks), the challenge is how to schedule and provision such functions on the heterogeneous resources efficiently. 
While various forms of heterogeneous-aware task scheduler already exist~\cite{salehi2016stochastic, denninnart2019improving, gentry2019robust} in the HPC context, serverless tasks are often of finer granularity. Therefore, a lightweight and low-latency scheduling that is aware of the machine heterogeneity is desired. 
 
 To make informed scheduling decisions in a heterogeneous system, a function execution-time profiler is needed to provide an estimated execution time of a given function on the heterogeneous machine types \cite{salehi2016stochastic}. Since each serverless function recurs multiple times, execution-time profiling can be carried out~\cite{shahrad2020serverless,wu2020descriptive}. However, to reduce the uncertainty in the execution-time prediction, thereby, making more informed scheduling decisions, on different machine types, function profiling must be performed proactively and in an explorative way to examine the infrequently-used options and collect their execution time statistics. Importantly, scheduling new user-defined functions, for which there is no prior execution-time statistics, is more challenging. Methods based on Transfer Learning \cite{torrey2010transfer} \revisedagain{(\ie reusing prior learning to speed up learning \finalrevise{a} new data set)} should be developed to enable inferring the execution time of the new function based on other existing functions on different machine types. A similar challenge and solution can be posed upon the addition of new machine types to the serverless system. 

\subsection{Performance Aspects in Serverless}

\subsubsection{Accelerator Multi-tenancy in Serverless Systems} 
Accelerator hardware is usually designed to be utilized by one tenant at a time. Even GPU is not originally designed to support multi-tenancy, although frameworks like Volta-MPS have added the multi-tenancy support to the recent models of GPU. Due to a serverless payment scheme where users do not pay for under-utilized hardware, it is cost-prohibitive to allocate the entire accelerator hardware to a single function. Maximizing the accelerator utilization implies deploying multiple functions sharing the same hardware \revisedagain{ while allowing authorized data sharing but not leaking sensitive data.} This is a logical development needed to enable economical accelerator deployment in serverless systems~\cite{risco2021gpu}.

\subsubsection{Task-Centric Scheduling} 
In the current commercial serverless computing platforms, functions are triggered without a specified deadline or urgency levels, thus, the scheduler treats tasks with equal priorities. Meanwhile, if task urgency and their demands are provided to the serverless platform scheduler, the task with low urgency can be executed in batch in favor of more serverless efficiency (see container reusing in Section~\ref{subsec:reuse_process}). Such urgency and demands for information are also helpful for the caching and other components to determine their operational priorities. Moreover, by detecting infeasible deadlines, the system can then promptly approximate the task, rather than missing its deadline because of exact computation. We envision a scheduler that looks into each task's QoS demand and then schedules them appropriately. However, requiring the user to profile their task's demand diminished the simplicity of the serverless platform. Another avenue of exploration is to identify the task urgency and their demands automatically at the platform level and transparently from the user's perspective.

\subsection{Edge-to-Cloud Serverless Platforms}

\subsubsection{Seamless Function Call \& Service Migration} 
The main benefits of serverless computing are the ease of use and abstracting of the users from the resource provisioning details. While the current serverless solutions are limited to cloud systems, it is desired to extend their benefits to the emerging edge-to-cloud systems and hide the complexity of dealing with multiple computing tiers (\ie device, edge, fog, and cloud tiers) from the user perspective.
The platform can transparently determine the appropriate tier for the function execution and can seamlessly migrate the execution from one tier to another to overcome the inherent resource scalability problem of the edge systems \cite{li2020heterogeneity}. These abilities can improve the system efficiency and unlock new use cases.
For instance, consider the use case of a blind person who uses smart glasses and needs real-time processing of the observed objects to enter a hotel lobby where few people are playing low-latency online games using the available edge system. Upon the arrival of the blind person, to make room for the blind application, the game functions have to be live-migrated to the cloud, so that the gaming is not interrupted. The opposite can happen when the blind person leaves the place. All these take place while the users have an illusion of everything being executed on their own devices. 
 
A serverless platform for the edge-to-cloud continuum extends the scope for computational reusability and approximation to circumstances where a task result that is already cached in another tier can be fetched from there, instead of executing it locally \cite{ffdn19}. Another interesting reusing potential that can be unleashed is in a scenario where edge devices forward common (reusable) tasks to a central cloud, so that other edge tiers can reuse them by fetching them from the cloud~\cite{satyanarayanan2019seminal}.

\subsubsection{Unified Management} 
From the service providers' perspective, it is a daunting task to manage the resources across cloud, fog, edge, and the device. Multi-tier applications are usually orchestrated to perform different duties on different tasks. For example, user devices take care of the UI/IO tasks, edge devices aggregate local data and then cloud machines do the heavy computing part. Such different tasks generally are written specifically to the hardware-specific tier.

What if all the applications---from a mobile device, edge to cloud devices---become serverless functions that run on the same standardized serverless framework that connects to the unified \revisedagain{mean} pane? Then, the function development, deployment, and migration can be done seamlessly within a single point of control. In addition, rather than each multi-tier system deploying its own edge devices separately, multiple multi-tier systems can also share the edge to cloud resources \revisedagain{and data}, since they all run on the same standardized framework. Such an idea can also be viewed as a CDN that not only provides static content at the edge but also allow serverless function executions on any local edge and fog server.

\subsection{Serverless Security and Trustworthiness}

\subsubsection{Trustworthiness}

Security is one of the criteria for developers in selecting a serverless platform \cite{kritikos2018review}.
With numerous components and the nature of function triggering, serverless computing exposes a large attack surface compared to its predecessors \cite{marin2021serverless}.
Poorly-designed functions meant for internal purposes often lack authentication and can be attacked via a direct triggering or an injection attack \cite{owasp}.
Meanwhile, the fact that serverless functions are stateless and short-lived limits the time available to attackers and the impact of successful attacks.
This shifts the attack strategies to become shorter and more indirect.

Cloud providers offer the shared responsibility model to the customers that make the cloud providers responsible for the infrastructure security and continuously applying security updates to the low-level software, hence, relieving some security burdens from the developers.
However, this leads to one common problem in cloud computing security: the trustworthiness of cloud providers \cite{spe17amini}. In the serverless paradigm, the trustworthiness issue is extended to cover several aspects: From the performance aspect, the developer cannot assure if the serverless platform achieves the required QoS, such as the number of available containers in the shared pool that are ready to serve the function's triggers;
From the isolation aspect, the developers cannot understand how the underlying isolation frameworks (VM/container) are configured for the functions. In particular, in serverless systems, the VMs can be potentially shared with other cloud users, due to the small function sizes. This increases the risk of being attacked.
Some serverless platforms \cite{awsserverlesscicd} offer built-in CI/CD pipelines to let developers conveniently build their source code to the function. This feature exposes other attack vectors, because the building steps may contain vulnerabilities (\eg the vulnerability of the third-party library) pulled by the platform to build the function \revisedagain{image. The topic of platform trustworthiness is highly critical to the adoption rate of data reusing techniques across users where the risk of data leak or compromise is a concern.}

\subsubsection{Secure Container}

As mentioned earlier, one of the security concerns in the serverless paradigm is the lack of isolation.
Containers expose vulnerabilities, such as privilege escalation from sharing the host kernel. To address the problem, the developers have choices to use rootless containers, or isolated containers \cite{rootlesscontainers} (\eg gVisor \cite{gvisor} or Kata Containers \cite{katacontainers}).
We note that, in practice, there should still be a trade-off between security and efficiency. The secure container offers high security but relatively low startup latency. On the other hand, relaxing isolation opens up more optimization opportunities \cite{sock216031}. Section~\ref{subsec:contreuse} discusses container instance reusing, which is an example of such relaxation.

\subsubsection{Denial of Wallet}

In the area of performance, serverless computing tolerates denial-of-service attacks more than its predecessors, because of its inherent ability to scale.
However, the ability to scale brings a new possibility for attackers to perform Denial-of-Wallet attacks  \cite{kelly2021denial}, an attack that forces the financial exhaustion of the application's owner instead of disabling the service availability. \finalrevise{When a service function is being targetted via a Denial-of-Wallet attack, the serverless platform steps in and triage the running cost down via enacting dropping and deferring low-importance tasks (see Section~\ref{subsec:schedlevel}) to preserve the user's funding and to keep the more important tasks running.}

\section{Summary}
\label{sec:conclsn}
It is envisioned that the future generation of highly scalable applications will predominantly rely on the serverless computing paradigm, hence, comprehensively studying the anatomy of this paradigm and identifying the scopes for efficiency can bring about major benefits to the users, developers, and providers. Accordingly, in this paper, we explored the ways to make the serverless computing paradigm efficient via investigating two main thrusts, namely computational reuse, and approximate computing. We started by characterizing the internal mechanics and studying different dimensions of serverless systems. Then, we surveyed the current state of the serverless and FaaS solutions (summarized in Table~\ref{tab:serverless_platformcmp}). Next, we categorized various approaches of reusing and approximation, respectively. An overview of these approaches is shown in Figure~\ref{fig:ReuseApprox}. In sum, we state that the characteristics of the serverless paradigm, where functions are compact, single-purpose, and portable, create a unique scope for efficiency, mostly via computational reuse, and then via approximate computing. 

In this paper, we also outlined several potential directions (summarized in Figure~\ref{fig:futurework}) that can push the envelope of the serverless paradigm toward the next generation of cloud computing systems. In summary, four prominent directions that we discussed as the future of the serverless paradigm are as follows: (A) Enabling higher-level abstractions, such as Object-as-a-Service and cloud-native programming languages, within the serverless paradigm; (B) Improving the performance of serverless clouds via exploiting the extensive function reusing and approximation opportunities exist in these systems; (C) Extending the serverless platforms beyond cloud infrastructure to cover multi-tier edge-to-cloud continuum, and (D) Extending the serverless abilities towards cloud-native serverless security.

\section*{Acknowledgments}
 We would like to thank the anonymous reviewers of the paper, and members of the HPCC Lab, particularly Davood G. Samani and Pawissanutt Lertpongrujikorn, who brainstormed with us on this paper. This research is supported by the National Science Foundation (NSF) under awards\# CNS-2047144 and CNS-2007209.

\newlength{\bibitemsep}\setlength{\bibitemsep}{.2\baselineskip plus .05\baselineskip minus .05\baselineskip}
\newlength{\bibparskip}\setlength{\bibparskip}{0pt}
\let\oldthebibliography\thebibliography
\renewcommand\thebibliography[1]{%
  \oldthebibliography{#1}%
  \setlength{\parskip}{\bibitemsep}%
  \setlength{\itemsep}{\bibparskip}%
}

\setlength{\bibitemsep}{.18\baselineskip plus .05\baselineskip minus .05\baselineskip}


\bibliographystyle{IEEEtran}

\bibliography{references}

\end{document}